%% file: main.tex
\documentclass[journal,10 pt,twocolumn]{IEEEtran}

\usepackage{graphicx}
\usepackage{balance}
\usepackage{comment}
\usepackage{cite}
\usepackage{amssymb}
\usepackage[tight,footnotesize]{subfigure}
\usepackage[active]{srcltx}
\usepackage{amsmath}
\usepackage{mathtools}

\graphicspath{{./figs/}}
\renewcommand{\baselinestretch}{0.95}
\usepackage{eurosym}
\usepackage[plain]{algorithm}
\usepackage{algorithmic}
\usepackage{multicol}
\usepackage{dsfont}
\usepackage{amsfonts}

\usepackage{cases}
\usepackage{xcolor}

\newlength\mylength 

\begin{document}

\title{Integrated Guidance and Gimbal Control for Coverage Planning with Visibility Constraints}

\author{Savvas~Papaioannou,~Panayiotis~Kolios,~Theocharis~Theocharides,\\~Christos~G.~Panayiotou~ and ~Marios~M.~Polycarpou
\thanks{The authors are with the KIOS Research and Innovation Centre of Excellence (KIOS CoE) and the Department of Electrical and Computer Engineering, University of Cyprus, Nicosia, 1678, Cyprus. {\tt\small \{papaioannou.savvas, pkolios, ttheocharides, christosp, mpolycar\}@ucy.ac.cy}}
}

\markboth{IEEE Transactions on Aerospace and Electronic Systems, doi:10.1109/TAES.2022.3199196, 2023. (EARLY ACCESS)}%
{Papaioannou \MakeLowercase{\textit{et al.}}: Integrated Guidance and Gimbal Control for Coverage Planning with Visibility Constraints}

\maketitle

\input{abstract}


\input{introduction}

\input{relatedWork}

\input{systemModel}

\input{approach}

\input{evaluation}

\input{conclusion}

\input{ack}

\bibliographystyle{IEEEtran}
\bibliography{main} 
\input{biography}

\flushbottom
\balance

\end{document}

%% file: abstract.tex
\begin{abstract}
Coverage path planning with unmanned aerial vehicles (UAVs) is a core task for many services and applications including search and rescue, precision agriculture, infrastructure inspection and surveillance. This work proposes an integrated guidance and gimbal control coverage path planning (CPP) approach, in which the mobility and gimbal inputs of an autonomous UAV agent are jointly controlled and optimized to achieve full coverage of a given object of interest, according to a specified set of optimality criteria. The proposed approach uses a set of visibility constraints to integrate the physical behavior of sensor signals (i.e., camera-rays) into the coverage planning process, thus generating optimized coverage trajectories that take into account which parts of the scene are visible through the agent's camera at any point in time. The integrated guidance and gimbal control CPP problem is posed in this work as a constrained optimal control problem which is then solved using mixed integer programming (MIP) optimization. Extensive numerical experiments demonstrate the effectiveness of the proposed approach. 
\end{abstract}

%% file: introduction.tex
\section{Introduction} \label{sec:Introduction}

\IEEEPARstart{O}{ver} the last years we have witnessed an accelerated demand for Unmanned Aerial Vehicles (UAVs) and Unmanned Aerial Systems (UASs) in various application domains including  search and rescue \cite{Moon2021,Papaioannou2019,Mishra2020,Papaioannou2020,Papaioannou2021a,Papaioannou2021b}, precision agriculture \cite{Mogili2018}, package delivery \cite{Das2020}, wildfire monitoring \cite{Afghah2019} and security \cite{PapaioannouJ1,PapaioannouJ2,Valianti2021}. This high demand is mainly fueled by the recent advancements in automation technology, avionics and intelligent systems, in combination with the proliferation and cost reduction of electronic components. 
Amongst the most crucial capabilities for a fully autonomous UAV system is that of path/trajectory planning \cite{Gasparetto2015}, which play a pivotal role in designing and executing automated flight plans, required by the majority of application scenarios. The path planning problem encapsulates algorithms that compute trajectories between the desired start and goal locations. Moreover, for many tasks such as structure inspection, target search, and surveillance, there is a great need for efficient and automated coverage path planning (CPP) \cite{Cabreira2019s} techniques. Coverage path planning consists of finding a path (or trajectory) which allows an autonomous agent (e.g., a UAV) to cover every point (i.e., the point must be included within the agent's sensor footprint) within a certain area of interest. 

Despite the overall technological progress in CPP techniques over the last decades, there is still work to be done for constructing solutions to the level of realism that can support practical autonomous UAV operations. As discussed in more detail in Sec. \ref{sec:Related_Work}, the vast majority of CPP approaches mainly consider ground vehicles or robots with static and fixed sensors (i.e., the sensor mounted on the robot is not controllable and the size of the sensor's footprint does not change). These assumptions reduce the CPP problem to a path-planning problem, where a) the area of interest is first decomposed into a finite number of cells (where usually each cell has size equal to the sensor's footprint), and b) the path that passes through all cells is generated with a path-finding algorithm, thus achieving full coverage. In some approaches (e.g., \cite{Xu2014}) the generated path is adapted to the robot's dynamic/kinematic model at a second stage. This two-stage approach however, usually produces sub-optimal results, in terms of coverage performance. In addition, UAVs are usually equipped with a gimballed sensor, which potentially exhibits a dynamic sensing range e.g., a pan-tilt-zoom (PTZ) camera. Therefore, in order to optimize coverage, necessitates the implementation of CPP techniques with the ability to optimize not only the UAV's trajectory, but also the control inputs of the onboard gimballed sensor.


Specifically, in this paper we investigate the coverage path planning problem for a known 2D region/object of interest, with a UAV agent, which exhibits a controllable gimbal sensor with dynamic sensing range. In particular, we propose an integrated guidance and gimbal control coverage path planning approach, where we jointly control and optimize a) the UAV's mobility input governed by its kinematic model and b) the UAV's gimbal sensor, in order to produce optimal coverage trajectories. The CPP problem is posed as a constrained optimal control problem, where the goal is to optimize a set of mission-specific objectives subject to coverage constraints. As opposed to the majority of existing techniques, in this work we consider the CPP problem in the presence of state-dependent visibility-constraints. In particular, the proposed approach integrates ray-casting into the planning process in order to determine which parts of the region/object of interest are visible through the UAV's sensor at any point in time, thus enabling the generation of realistic UAV trajectories.

Specifically, the contributions of this work are the following:

\begin{itemize}
    \item We propose an integrated guidance and gimbal control CPP approach for the problem of coverage planning in the presence of kinematic and sensing constraints including state-dependent visibility constraints i.e., we are simulating the physical behavior of sensor signals in order to account for the parts of the scene that are blocked by obstacles and thus cannot be observed by the UAV's sensor at a given pose.
    \item We formulate the CPP problem as a constrained optimal control problem, in which the UAV mobility and gimbal inputs are jointly controlled and optimized according to a specified set of optimality criteria, and we solve it using  mixed integer programming (MIP). The performance of the proposed approach is demonstrated through a series of numerical experiments.
\end{itemize}

The rest of the paper is organized as follows. Section~\ref{sec:Related_Work} summarizes the related work on coverage path planning with ground and aerial vehicles. Then, Section \ref{sec:system_model} develops the system model, Section \ref{sec:problem} defines the problem tackled and Section \ref{sec:approach} discusses the details of the proposed coverage planning approach. Finally, Section \ref{sec:Evaluation} evaluates the proposed approach and Section \ref{sec:conclusion} concludes the paper and discusses future work.

%% file: relatedWork.tex
\section{Related Work}\label{sec:Related_Work}

In coverage path planning (CPP) we are interested in determining the path/trajectory that enables an autonomous agent to observe with its sensor every point in a given environment. Early works e.g., \cite{Choset1998,Acar2002}, treated the CPP problem as a path-planning problem, by decomposing the environment into several non-overlapping cells, and then employing a path-planning algorithm \cite{Nadhir2020} to find the path that passes through every cell. Notably, in \cite{Choset1998,Acar2002} the authors propose cellular decomposition coverage algorithms, where the free-space of the environment is decomposed into non-intersecting regions or cells, which can be covered by the robot, one by one using simple back-and-forth motions, thus covering the whole area. Extensions \cite{Acar2006,Huang2001,Mnnadiar2010} of this approach have investigated the coverage/sweeping pattern, the sensor's footprint and the region traversal order for optimal coverage. In \cite{Gabriely2001}, the authors propose a CPP approach based on spanning trees, for covering, in linear time, a continuous planar area with a mobile robot equipped with a square-shaped sensor footprint. In the subsequent work \cite{Gabriely2002}, the plannar area to be covered is incrementally sub-divided into disjoint cells on-line, allowing for real-time operation. The spanning tree approach presented in \cite{Gabriely2001} and \cite{Gabriely2002} is extended in \cite{Agmon2008} for multi-robot systems. In \cite{Karapetyan2017}, the boustrophedon cellular decomposition algorithm (BDC) \cite{Choset2000} is used to compute a graph-based representation of the environment, and then a coverage clustering algorithm is proposed which divides the coverage path among multiple robots. The problem of coverage path planning has also been investigated with multiple ground robots in \cite{Janchiv2013}. The authors of \cite{Janchiv2013} propose an approach that minimizes the total coverage time by transforming the CPP problem into a network flow problem. In \cite{Cannata2011}, a graph-based coverage algorithm is proposed for a CPP problem variation, which enables a team of robots to visit a set of predefined locations according to a specified frequency distribution. Another CPP variation is investigated in \cite{Schwager2009,Cortes2004}. The authors propose decentralized, adaptive control laws to drive a team of mobile robots to an optimal sensing configuration which achieves maximum coverage of the area of interest. The same problem is investigated in \cite{Pimenta2008} with heterogeneous robots that exhibit different sensor footprints.

Interestingly, the CPP problem has recently gained significant attention due to its importance in various UAV-based applications, including emergency response \cite{Savkin2020,Papaioannou2021a,Papaioannou2021b}, critical infrastructure inspection \cite{Phung2019} and surveillance \cite{Savkin2019}. A UAV-based coverage path planning approach utilizing exact cellular decomposition in polygonal convex areas is proposed in \cite{Li2011}, whereas in \cite{Xu2014} the CPP problem is adapted for a fixed-wing UAV. In \cite{Xu2014} the ground plane is first decomposed into several non-overlapping cells which are then connected together forming a piecewise-linear coverage path. At a second stage a UAV-specific motion controller is used to convert the generated path into a smooth trajectory which the UAV can execute. An information theoretic coverage path planning approach for a single fixed-wing UAV is presented in \cite{Liam2014}. The aircraft maintains a coverage map which uses to it make optimized control decisions on-line, in order to achieve global coverage of the environment. In \cite{Chen2021}, the authors propose a clustering-based CPP approach for searching multiple regions of interest with multiple heterogeneous UAVs, by classifying the various regions into clusters and assigning the clusters to the UAVs according to their capabilities. In \cite{Bentz2018,Cabreira2019,Choi2020} CPP techniques for energy-constraint multi-UAV systems are proposed. Moreover, in \cite{Elmokadem2019}, a distributed coverage control approach is proposed for a multi-UAV system. Specifically, a coverage reference trajectory is computed using a Voronoi-based partitioning scheme, and then a distributed control law is proposed for guiding the UAVs to follow the reference trajectory, thus achieving full coverage. Finally, the CPP problem has also been investigated more recently with learning based techniques \cite{Theile2020,Sanna2021}. In \cite{Theile2020}, an end-to-end deep reinforcement learning CPP approach is proposed for an autonomous UAV agent. The authors utilize a double deep Q-network to learn a coverage control policy that takes into account the UAV's battery constraints. On the other hand in \cite{Sanna2021}, a coverage path planning system based on supervised imitation learning is proposed for a team of UAV agents. This approach plans coordinated coverage trajectories which allow unexplored cells in the environment to be visited by at least one UAV agent. 

The problem of coverage path planning (CPP) is also related to the view planning problem (VPP) and its variations, where the objective is to find the minimum number of viewpoints that completely cover the area of an object of interest. The relationship between these two problems is showcased in \cite{R1}, in which the CPP problem is posed as a view planning problem. Specifically, the authors in \cite{R1} propose an algorithm based on the traveling salesman problem (TSP), which incorporates visibility constraints and finds the tour of minimum length which allows a UAV equipped with a fixed downward facing camera to inspect a finite sets of points on the terrain. On the other hand, the authors in \cite{R2} propose a variation of the original VPP problem, termed traveling VPP, where the objective is the minimization of the combined view and traveling cost i.e., the cost to minimize combines the view cost which is proportional to the number of viewpoints planned, and the traveling cost which accounts for the total distance that the robot needed to travel in order to cover all points of interest. In general, VPP approaches operate in a discrete state-space setting and are mostly concerned with the selection of an optimal sequence of sensing actions (taken from the finite set of all admissible actions) which achieve full coverage of the object of interest. In contrast, the proposed approach can be used to generate continuous trajectories which are governed by kinematic and sensing constraints. The interested reader is directed to \cite{Galceran2013,Cabreira2019s,Scott2003} for a more detailed examination of the various coverage path planning techniques in the literature, including the view planning problem.

To summarize, in comparison with the existing techniques, in this work we propose a coverage planning approach which integrates ray-casting into the planning process, in order to simulate the physical behavior of sensor signals and thus determine which parts of the scene are visible through the UAV's onboard camera. The proposed approach takes into account both the kinematic and the sensing constraints of the UAV agent, to achieve full coverage of an object of interest in the presence of obstacles. In particular, the coverage path planning problem is posed in this work as a constrained open-loop optimal control problem which incorporates the UAV's kinematic and sensing model to produce optimal coverage trajectories in accordance with the specified mission objectives. Finally, the proposed mathematical formulation can be solved optimally with off-the-shelf mixed integer programming (MIP) optimization tools \cite{Anand2017}.

%% file: systemModel.tex
\section{System Model} \label{sec:system_model}

\subsection{Agent Kinematic Model} \label{ssec:kinematic_model}

This work assumes that an autonomous agent (e.g., a UAV), is represented as a point-mass object which maneuvers inside a bounded surveillance area $\mathcal{W} \subset \mathbb{R}^2$. The agent kinematics are governed by the following discrete-time linear model \cite{Luis2020}:
\begin{equation} \label{eq:agent_dynamics}
    x_{t} = \Phi x_{t-1} + \Gamma u_{t-1}, 
\end{equation}
where $x_t = [p_t(x),p_t(y),\nu_t(x),\nu_t(y)]^\top$ denotes the agent's state at time $t$ in cartesian $(x, y)$ coordinates, which consists of the agent's position $[p_t(x),p_t(y)]^\top \in \mathbb{R}^2$ and velocity  $[\nu_t(x),\nu_t(y)]^\top \in \mathbb{R}^2$ in the $x$ and $y$ directions. The term $u_{t} = [f_t(x), f_t(y)]^\top \in \mathbb{R}^2$ denotes the control input, i.e., the amount of force applied in each dimension in order to direct the agent according to the mission objectives. 
The matrices $\Phi$ and $\Gamma$ are given by:
\begin{equation}
\Phi = 
\begin{bmatrix}
    \text{I}_{2\times2} & \delta t \cdot \text{I}_{2\times2}\\
    \text{0}_{2\times2} & \phi \cdot \text{I}_{2\times2}
   \end{bmatrix},~
\Gamma = 
\begin{bmatrix}
    \text{0}_{2\times2} \\
     \gamma \cdot \text{I}_{2\times2}
   \end{bmatrix},
\end{equation}

\noindent where $\delta t$ is the sampling interval, $\text{I}_{2\times2}$ and $\text{0}_{2\times2}$ are the 2 by 2 identity matrix and zero matrix respectively, with $\phi$ and $\gamma$ given by $\phi =  (1-\eta)$ and $\gamma = m^{-1} \delta t$, where the parameter $\eta$ is used to model the (air) drag coefficient and $m$ is the agent mass. Given a known initial agent state $x_0$, and a set of control inputs $\{u_t | t=0,..,T-1\}$, inside a finite planning horizon of length $T$, the agent trajectory can be obtained for time-steps $t=[1,..,T]$ by the recursive application of Eqn. \eqref{eq:agent_dynamics} as:

\begin{equation}\label{eq:unrolled_kinematics}
    x_{t} = \Phi^{t} x_{0} + \sum_{\tau=0}^{t-1} \Phi^{t-\tau-1} \Gamma u_{\tau}.
\end{equation}

\noindent Therefore, the agent's trajectory can be designed and optimized to meet the desired mission objectives by appropriately selecting the control inputs $\{u_t | t=0,..,T-1\}$, inside  the given planning horizon of length $T$. We should point out here that although the agent kinematic model in Eqn. \eqref{eq:agent_dynamics} does not fully captures the low-level UAV aerodynamics (which are platform dependent), it allows us to design and construct high-level (i.e., mission-level) coverage trajectories which in turn can be used as desired reference trajectories to be tracked with low-level closed-loop guidance controllers found on-board the UAVs\cite{garcia2011nonlinear,gavilan2015iterative,Cowling2007prototype}.

\subsection{Agent Sensing Model} \label{ssec:sensing_model}

The agent is equipped with a gimbaled camera with optical zoom, which is used for sensing its surroundings and performing various tasks e.g., searching objects/regions of interest, detecting targets, etc. The camera field-of-view (FoV) or sensing footprint is modeled in this work as an isosceles triangle \cite{Wang2021visibility,Penin2018vision} parameterized by its angle at the apex $\varphi$ and its height $h$, which are used to model the FoV angle opening and sensing range respectively. We should point out here that any convex 2D shape can be used to model the camera FoV. \color{black} Using the parameters $\varphi$ and $h$ the camera FoV side length $(\ell_s)$ and base length $(\ell_b)$ are computed as:
\begin{equation}
    \ell_s = h \times \text{cos}(\varphi/2)^{-1}, ~\text{and},~ \ell_b =2 \ell_s \times \text{sin}(\varphi/2).
\end{equation}

\noindent Therefore, the set of vertices $(\mathcal{V}_o)$ of the triangular FoV camera projection, for an agent centered at the origin, and facing downwards are given by $\mathcal{V}_o=[v_1,v_2,v_3]$, where $v_1 = [0, 0]^\top$, $v_2 = [-\ell_b/2, -h]^\top$ and $v_3 = [\ell_b/2, -h]^\top$ so that:

\begin{equation}
    \mathcal{V}_o =
    \begin{bmatrix}
       0 & -\ell_b/2 & \ell_b/2 \\
       0 & -h & -h \\
    \end{bmatrix},
\end{equation}

The camera FoV can be rotated (on the $xy$-plane) around the agent's position $x^\text{pos}=[p(x),p(y)]^\top$ by an angle $\theta \in \bar{\Theta}$ (with respect to $x$-axis), by performing a geometric transformation consisting of a rotation operation followed by a translation:
\begin{equation}\label{eq:fov_vertices}
    \mathcal{V} = R(\theta)\mathcal{V}_o + x^\text{pos},
\end{equation}

\noindent where $\mathcal{V}$ is the rotated camera FoV in terms of its vertices, $\theta$ is the control signal and $R(\theta)$ is a 2D rotation matrix given by:
\begin{equation} \label{eq:rotation_mat}
   R(\theta) =
    \begin{bmatrix}
       \text{cos}(\theta) & \text{sin}(\theta) \\
       -\text{sin}(\theta) & \text{cos}(\theta) 
    \end{bmatrix}.
\end{equation}

\begin{figure}
	\centering
	\includegraphics[width=\columnwidth]{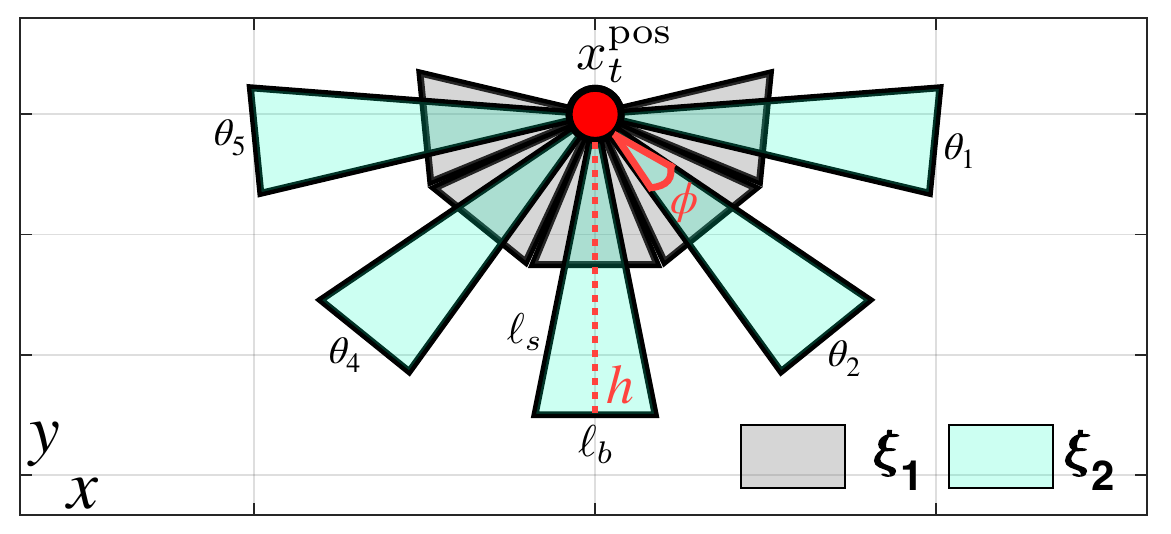}
	\caption{The figure illustrates all the possible sensor FoV configurations for two different zoom-levels $\xi_1$ and $\xi_2$, and for 5 different rotation angles $\theta_1,..,\theta_5$, when the agent's position is equal to $x_t^\text{pos}$. In this example, the total number of FoV configurations is equal to $|\bar{\Theta}|\times|\bar{\Xi}|=10$.}	
	\label{fig:fig1}
	\vspace{-0mm}
\end{figure}

\noindent We should mention here that in this work the rotation angle $\theta$ takes its values from a finite set of all admissible rotation angles $\bar{\Theta} = \{\theta_1,..,\theta_{|\bar{\Theta}|}\}$, where $|\bar{\Theta}|$ denotes the set cardinality. The agent can be placed in any desired position and orientation (i.e., pose) at some time-step $t$, by adjusting the control signals $u_t$ and $\theta_t \in |\bar{\Theta}|$ i.e., $\mathcal{V}_t = R(\theta_t)\mathcal{V}_o + x^\text{pos}_t$.
The agent's onboard camera is also equipped with an optical zoom functionality, which can alter the FoV characteristics in order to better suit the mission objectives and constraints. In particular, it is assumed that a zoom-in operation narrows down the FoV (i.e., reduces the FoV angle opening $\varphi$) however, it increases the sensing range $h$, as shown in Fig. \ref{fig:fig1}. \color{black} In particular, we assume that the camera exhibits a finite set of predefined zoom levels denoted as $\bar{\Xi}=\{\xi_1,...,\xi_{|\bar{\Xi}|}|\xi_i \in \mathbb{R}, \xi_i \ge 1\}$, which are used to scale the FoV parameters. The zoon-in functionality is thus defined for a particular zoom level $\xi \in \bar{\Xi}$ as:
\begin{equation}
    h^\prime = h\times \xi, ~\text{and},~\varphi^\prime = \varphi \times \xi^{-1},
\end{equation}


\noindent where $(h^\prime,\varphi^\prime)$ are the new parameters for the FoV angle and range, after applying the optical zoom level $\xi$. A visual representation of the camera model is illustrated in Fig. \ref{fig:fig1}.

We can now denote the agent's sensing state at time-step $t$ as $\mathcal{S}_t(\theta_t,x^\text{pos}_t,\xi_t)$, which jointly accounts for the agent's position and orientation. The notation $\mathcal{S}_t(\theta_t,x^\text{pos}_t,\xi_t)$, is used here to denote the area inside the agent's sensing range determined by the FoV vertices (i.e., the convex hull) as computed by Eqn. \eqref{eq:fov_vertices}. Therefore, the total area covered by the agent's FoV within a planning horizon of length $T$ can thus be obtained by:
\begin{equation}\label{eq:total_fov}
    \mathcal{S}_{1:T} = \bigcup_{t=1}^T ~\mathcal{S}_t(\theta_t,x^\text{pos}_t,\xi_t),
\end{equation}
\noindent where  $\theta_t \in \bar{\Theta}$, $\xi_t \in \bar{\Xi}$ and the agent position $x^\text{pos}_t$ has been computed from the application of a set of mobility controls inputs $\{u_\tau| \tau=0,..,t-1\}$ according to Eqn. \eqref{eq:unrolled_kinematics}. 

We should point out here that the agent kinematic and sensing models described above can easily be extended to 3 dimensions (i.e., the agent kinematics in Eqn. \eqref{eq:agent_dynamics} can be extended in 3D by accounting for the $z$ dimension, the triangular 2D FoV translates to a regular pyramid in 3D, the 2D rotation matrix $R$ becomes a 3D rotation matrix in 3D,  etc). Consequently, the proposed approach discussed in detail in Sec.\ref{sec:approach} can also be extended in 3D environments with some modifications. However, in order to make the analysis of the proposed approach easier to follow and more intuitive, the problem in this paper has been formulated in a two dimensional space, which already has some key challenges.
\color{black}

\section{Problem Statement} \label{sec:problem}
Let an arbitrary bounded convex planar region $\mathcal{C} \subset \mathcal{W}$ to denote a single object or region of interest, that we wish to cover with our autonomous agent, with boundary $\partial\mathcal{C}$, as shown in Fig. \ref{fig:fig2}. The proposed approach can be used to generate the coverage plan for the area enclosed in $\mathcal{C}$ in the case where the region $\mathcal{C}$ is traversable. On the other hand when  $\mathcal{C}$ is not traversable (i.e., $\mathcal{C}$ represents an inaccessible, to the agent, object or region of interest), the proposed technique is used to generate the coverage plan for the region's boundary $\partial\mathcal{C}$. For brevity, we will formulate the problem assuming the latter scenario (i.e., generating coverage plans for covering the boundary of a region/object of interest), however the proposed formulation can be applied for both scenarios.
In a high level form, the problem tackled in this work can be formulated as follows: 

\begin{algorithm}
\begin{subequations}
\begin{align}
&\hspace*{-5mm}\textbf{Problem (P1):}~\texttt{High-level Controller} &  \nonumber\\
& \hspace*{-5mm}~~~~\underset{\mathbf{U}_T, \mathbf{\Theta}_T, \mathbf{\Xi}_T}{\arg \min} ~\mathcal{J}_\text{coverage}(\mathbf{X}_T, \mathbf{U}_T, \mathbf{\Theta}_T, \mathbf{\Xi}_T) &\label{eq:objective_P1} \\
&\hspace*{-5mm}\textbf{subject to: $t=[1,..,T]$} ~  &\nonumber\\
&\hspace*{-5mm}  x_{t} = \Phi^{t} x_{0} + \sum_{\tau=0}^{t-1} \Phi^{t-\tau-1} \Gamma u_{\tau} & \label{eq:P1_1}\\
&\hspace*{-5mm} x_0, x^\text{pos}_{t} \notin \psi,~ \forall \psi \in \Psi & \label{eq:P1_2}\\
&\hspace*{-5mm} \partial\mathcal{C} \in \mathcal{S}^\prime_{1:T} & \label{eq:P1_3}\\
&\hspace*{-5mm} x_0, x_{t} \in \mathcal{X}, ~ u_t \in \mathcal{U},~ \theta_{t} \in \bar{\Theta},~ \xi_{t} \in \bar{\Xi}  & \label{eq:P1_4}
\end{align}
\end{subequations}
\end{algorithm}

In (P1) we are interested in finding the agent's mobility (i.e., $\mathbf{U}_T=\{u_0,..,u_{T-1}\}$) and sensor  (i.e., $\mathbf{\Theta}_T=\{\theta_1,..,\theta_T\}$, $\mathbf{\Xi}_T = \{\xi_1,..,\xi_T\}$) control inputs, over the planning horizon of length $T$, i.e.,  $\mathcal{T} = \{1,..,T\}$, which optimize a certain set of optimality criteria encoded in the state-dependent multi-objective cost function $\mathcal{J}_\text{coverage}(\mathbf{X}_T, \mathbf{U}_T, \mathbf{\Theta}_T, \mathbf{\Xi}_T)$, where $\mathbf{X}_T=\{x_1,..,x_T\}$, subject to the set of constrains shown in Eqn. \eqref{eq:P1_1}-\eqref{eq:P1_4}. In particular, Eqn. \eqref{eq:P1_1} represents the agent's kinematic constraints as described in Sec. \ref{ssec:kinematic_model}. Then, Eqn. \eqref{eq:P1_2} represents obstacle avoidance constraints with a specified set of obstacles $\Psi$, and the constraint in \eqref{eq:P1_3} (i.e., coverage constraint) is used to guarantee that during the planning horizon the whole boundary of the region/object of interest will be covered by the agent's sensor FoV.

We should mention here that the notation $\mathcal{S}^\prime_{1:T} \subseteq \mathcal{S}_{1:T}$ refers to the reduced FoV coverage obtained when obstacles block the sensor signals (i.e., camera-rays) from passing through, thus creating occlusions. In this work we use a set of visibility constraints to distinguish between parts of the scene $p \subset \partial\mathcal{C}$ that belong to the visible field-of-view i.e., $p \in \mathcal{S}^\prime_{1:T}$ and parts $p$ that are occluded. In order to model the visible field-of-view $\mathcal{S}^\prime_{1:T}$ we use ray-casting to simulate the physical behavior of the camera-rays and account for the occluded regions. Therefore, the constraint in Eqn. \eqref{eq:P1_3}, enforces the generation of coverage trajectories, which take into account which parts of the scene are visible through the agent's camera at any point in time. 
Finally, the constraints in Eqn. \eqref{eq:P1_4} restrict the agent's state and control inputs within the desired bounds. In the next section, we discuss in more detail how we have tackled the problem discussed above.

We should mention here that in this work the following assumptions are made: a) the agent has self-localization capability (e.g., via accurate GPS positioning), b) the environment (i.e., object of interest, obstacles, etc.) is known a-priori, and c) the visual data acquisition process is noise-free. However, in certain scenarios in which the assumptions above no longer apply, the agent's visual localization accuracy at the planning time is of essence in generating optimal coverage trajectories. In such scenarios the proposed approach can be combined with active visual localization techniques \cite{Zhang2020fisher}, in order to improve the agent's visual localizability, and generate accurate coverage trajectories.

%% file: approach.tex
\section{Integrated Guidance and Gimbal Control Coverage Planning}\label{sec:approach}
In this section we design a mixed integer quadratic program (MIQP) in order to tackle the optimal control problem of integrated guidance and gimbal control coverage planning, as described in problem (P1).

\subsection{Preliminaries}\label{ssec:preliminaries}
\begin{figure}
	\centering
	\includegraphics[width=\columnwidth]{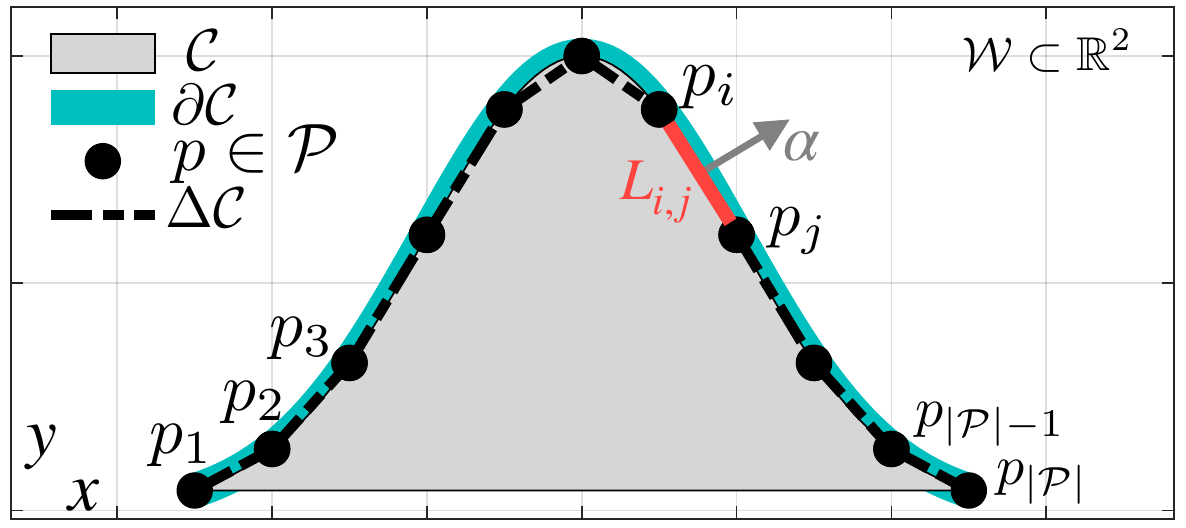}
	\caption{In the scenario above we are interested in finding the optimal mobility and sensor control inputs for a single UAV agent, which enable full coverage of the boundary $\partial \mathcal{C}$ of a given region of interest $\mathcal{C}$. $\Delta \mathcal{C}$ denotes the  piece-wise linear approximation of the boundary which is composed of a finite number of line segments. The points on the boundary $p_i$ and $p_j$ are connected with the line segment $L_{i,j} \in \Delta \mathcal{C}$, where $\alpha$ denotes its outward normal vector. Full coverage is achieved when every point $p \in \mathcal{P}$ is included inside the agent's sensor FoV.}	
	\label{fig:fig2}
	\vspace{-0mm}
\end{figure}

The proposed approach first proceeds by sampling points $p \in \partial \mathcal{C}$ on the region's boundary, generating the set of points $\mathcal{P} = \{p_1,..,p_{|\mathcal{P}|}\} \subset \partial \mathcal{C}$, where $|\mathcal{P}|$ is the set cardinality, thus creating a discrete representation of the region's boundary that needs to be covered, as depicted in Fig. \ref{fig:fig2}. Equivalently, a set of points $\mathcal{P} \subset \mathcal{C}$ are sampled from $\mathcal{C}$ in the scenario where the region of interest is traversable. Essentially, the object of interest is represented in this work as a point-cloud, which in practice can be obtained with a number of scene/object reconstruction techniques \cite{Guo2020deep,Newcombe2010}. The coverage constraint shown in Eqn. \eqref{eq:P1_3} can now be written as:
\begin{equation}\label{eq:discrete_coverage_con}
     p \in \mathcal{S}^\prime_{1:T}, ~\forall p \in \mathcal{P},
\end{equation}

\noindent and thus we are looking to find the optimal agent mobility and sensor control inputs over the planning horizon, which satisfy the constraint in Eqn. \eqref{eq:discrete_coverage_con}, i.e., the set of points $\mathcal{P}$ must be covered by the agent's sensor FoV. We should mention here that the methodology used to generate $\mathcal{P}$ (e.g., systematic selection or sampling procedure) is up to the designer and in accordance to the problem requirements. 

In the special scenario examined in this work, in which the region (or object) of interest $\mathcal{C}$ is not traversable, and thus acting as an obstacle to the agent's trajectory, a set of obstacle avoidance constraints are implemented i.e., Eqn. \eqref{eq:P1_2} to prevent collisions between the agent and the obstacle. Intuitively, in such scenarios the agent must avoid crossing the region's boundary $\partial \mathcal{C}$. We denote as $\Delta \mathcal{C}$, the piece-wise linear approximation of the boundary $\partial \mathcal{C}$, that contains the line segments $L_{p_i}=L_{i,j} = \{p_i+ r (p_j-p_i)| r \in [0,1], j \ne i\}$, which are formed when connecting together the pair of points $(p_i,p_j)_{i \ne j} \in \mathcal{P}$, such that the resulting line segments $L_{i,j}$ belong to the boundary of the convex hull of $\mathcal{P}$, as shown in Fig. \ref{fig:fig2}. 

Let us assume that $|\mathcal{P}|$ points $\{p_1,..,p_{|\mathcal{P}|}\}$ have been sampled from $\partial \mathcal{C}$, and that $\Delta \mathcal{C}$ contains $|\mathcal{P}|$ line segments $\{L_{p_1},..,L_{p_{|\mathcal{P}|}}\}$, where each line segment $L_{p_i}, i=1,..,|\mathcal{P}|$ lies on the line $\tilde{L}_{p_i} = \{x \in \mathbb{R}^{(2,1)} | \alpha^\top_i x= \beta_i \}$, where the line coefficients in the vector $\alpha_i$ determine the outward normal to the $i_\text{th}$ line segment and $\beta_i$ is a constant. The area inside the region of interest $\mathcal{C}$ is thus modeled as a convex polygonal obstacle with boundary $\Delta \mathcal{C}$ defined by $|\mathcal{P}|$ linear equations $\tilde{L}_{p_i}, i=1,..,|\mathcal{P}|$. A collision occurs when at some point in time $t \in \mathcal{T}$, the agent's position $x_t^\text{pos}$ resides within the area defined by the region's boundary $\Delta \mathcal{C}$ or equivalently the following system of linear inequalities is satisfied: 
\begin{equation} \label{eq:colision1}
    \alpha^\top_i x_t^\text{pos} < \beta_i,~ \forall i \in \left[1,..,|\mathcal{P}|\right].
\end{equation}

\noindent Hence, a collision can be avoided with $\mathcal{C}$ at time $t$ i.e., $x_t^\text{pos} \notin \mathcal{C}$,  when $\exists~ i \in [1,..,|\mathcal{P}|]: \alpha^\top_{i} x_t^\text{pos} \ge  \beta_i$. This is implemented with a set of mixed integer linear constraints as shown in Eqn. \eqref{eq:O_1}-\eqref{eq:O_3}.

\begin{algorithm}
\begin{subequations}
\begin{align}
& -\alpha^\top_{i} x_t^\text{pos} - M b^\text{collision}_{t,i} \le -\beta_{i},~\forall t,i, \label{eq:O_1}\\
& \sum_{i=1}^{|\mathcal{P}|} b^\text{collision}_{t,i} \le  (|\mathcal{P}|-1), ~ \forall t, \label{eq:O_2} \\
& b^\text{collision}_{t,i} \in \{0,1\}, ~ \forall t, i. \label{eq:O_3}
\end{align}
\end{subequations}
\end{algorithm}

\noindent Specifically, the constraint in Eqn. \eqref{eq:O_1}, uses the binary variable $b^\text{collision}_{t,i}$ to determine whether the $i_\text{th}$ inequality i.e., $\alpha^\top_{i} x_t^\text{pos} <  \beta_i$ of Eqn. \eqref{eq:colision1} is satisfied at some time $t \in \mathcal{T}$ by setting $b^\text{collision}_{t,i}=1$, where $M$ is a large positive constant.  Then, the constraint in Eqn. \eqref{eq:O_2} makes sure that for any time-step $t$, the binary variable $b^\text{collision}_{t,i}$ is activated less than $|\mathcal{P}|-1$ times, to ensure that the agent's position $x_t^\text{pos}$ does not reside inside the obstacle. These constraints, can be applied for any number of convex polygonal obstacles $\psi \in \Psi$, that need to be avoided, by augmenting the variable $b^\text{collision}_{t,i}$ with an additional index to indicate the obstacle number.

\subsection{Visibility Constraints} \label{ssec:vis_con}
In the previous section we have described how the region's boundary $\partial \mathcal{C}$ that needs to be covered, is defined as a piece-wise linear approximation $\Delta \mathcal{C}$, and we have also shown how the non-traversable area inside the region of interest $\mathcal{C}$, is modeled with a set of obstacle avoidance constraints, which are used to prevent the agent from passing-through. In this section, we devise a set of visibility constrains, which allows us to determine which parts of the region's boundary are visible (i.e., not blocked by an obstacle) through the agent's camera, given a certain agent pose.

In this work, we use the term camera-rays to denote the light rays that are captured by the camera's optical sensor. Without loss of generality, let us assume that at each time-step $t$, a finite set of camera-rays enter the optical axis, denoted as $\mathcal{K}_{\theta_t,x^\text{pos}_t,\xi_t}=\{K_1,..,K_{|\mathcal{K}|}\}$, where $\theta_t,x^\text{pos}_t,\xi_t$ determine the agent's pose and subsequently the FoV configuration, as illustrated in Fig. \ref{fig:fig3}(a)-(b).

 The individual ray $K_i$ is given by the line-segment $K_i = \{x_t^\text{pos} + s(\kappa_i-x_t^\text{pos}) | s \in [0,1]\}$, where $x_t^\text{pos}$ is the ray's origin given by the agent's position at time $t$ and $\kappa_i \in \mathbb{R}^2$ is a fixed point on the base of the triangle which defines the camera FoV and determines the ray's end point. We can now define the notion of visibility as follows: The point $p_i \in \mathcal{P}, i \in [1,..,|\mathcal{P}|]$ on the region's boundary $\partial \mathcal{C}$, which exists on the line-segment $L_{p_i} \in \Delta \mathcal{C}$ is said to be visible at time-step $t$ i.e., $p_i \in \mathcal{S}^\prime_t(\theta_t,x^\text{pos}_t,\xi_t)$ when: 
\begin{equation}\label{eq:1_1}
    p_i \in \mathcal{S}_t(\theta_t,x^\text{pos}_t,\xi_t)  ~\wedge~ \exists K \in \mathcal{K}_{\theta_t,x^\text{pos}_t,\xi_t} : (K \otimes \Delta \mathcal{C}) = L_{p_i},
\end{equation}

\noindent where the operation $K \otimes \Delta \mathcal{C}$ is defined as the intersection of the camera-ray $K$, with the set of line segments in $\Delta \mathcal{C}$, and returns the nearest (i.e., closest distance with respect to the ray's origin) line-segment $L \in \Delta \mathcal{C}$ which the ray $K$ intersects with. In the case where a ray $K$ exhibits no intersections with any line-segment, $\emptyset$ is returned instead. In essence, the point $p_i$ is visible at time $t$, when both constraints in Eqn. \eqref{eq:1_1} are satisfied i.e.,  a) $p_i$ is included inside the agent's camera FoV $\mathcal{S}_t(\theta_t,x^\text{pos}_t,\xi_t)$ and b) there exists a camera-ray $K$ which first intersects with the line-segment $L_{p_i}$ which contains the point $p_i$. In other words there is a camera-ray $K$ which does not intersect with any parts of the boundary (i.e., line segments) $L \in \Delta \mathcal{C}$ prior to $L_{p_i}$, as illustrated in Fig. \ref{fig:fig3}(b).

Let the camera-ray $K$ (i.e., $K = \{x_t^\text{pos} + s(\kappa-x_t^\text{pos}) | s \in [0,1]\}$) to have  $x$ and $y$ cartesian coordinates given by $x_t^\text{pos}(x) + s[\kappa(x) - x_t^\text{pos}(x)]$ and $x_t^\text{pos}(y) + s[\kappa(y) - x_t^\text{pos}(y)]$ respectively. Also, let the $x$ and $y$ cartesian coordinates of the a line-segment $L_{p_i}=L_{i,j} \in \Delta \mathcal{C}$ on the boundary (i.e., $L_{i,j} = \{p_i+ r (p_j-p_i)| r \in [0,1], j \ne i\}$) to be $p_i(x) + r[p_j(x) - p_i(x)]$ and $p_i(y) + r[p_j(y) - p_i(y)]$ respectively. The intersection $K \otimes L_{i,j}$ of the camera-ray with the line segment  can be computed by solving the following system of linear equations for the two unknowns i.e., $(s,r)$:
\begin{figure*}
	\centering
	\includegraphics[width=\textwidth]{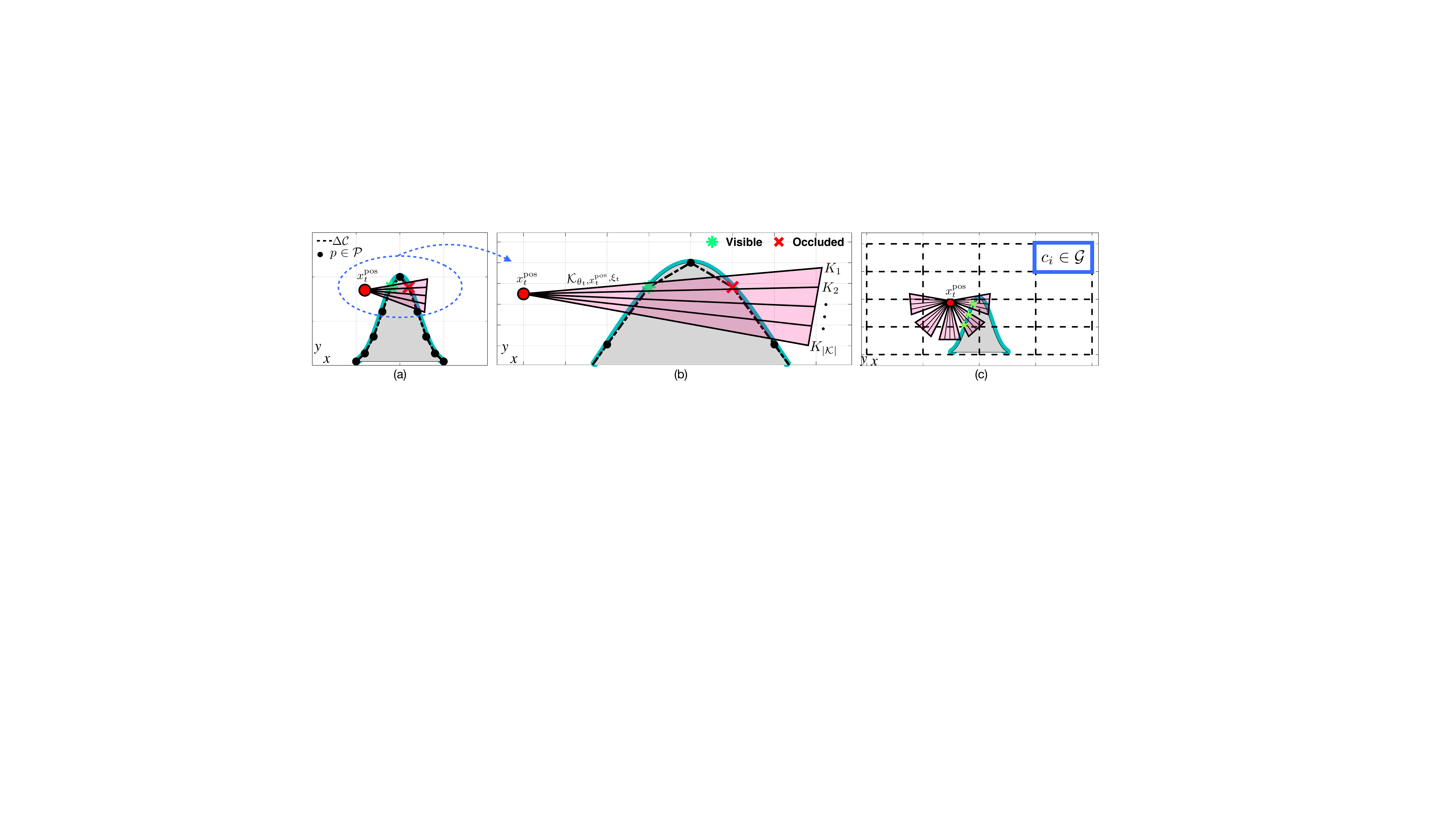}
	\caption{The figure illustrates how the proposed approach tackles the visibility problem by simulating the physical behavior of camera rays. (a) In the coverage problem investigated in this work, every point $p \in \mathcal{P}$ on the boundary approximation $\Delta \mathcal{C}$ of the region of interest must be covered by the agent's sensor FoV, shown with the pink triangle. (b) The sensor's FoV is composed of a finite number of camera-rays $K_1,..,K_{|\mathcal{K}|} \in \mathcal{K}_{\theta_t,x^\text{pos}_t,\xi_t}$ parameterized by the agent's pose as shown above. A point $p$ is visible through the agent's camera when it resides inside the camera FoV and exists some ray $K$ which first intersects the line segment $L_p$ which contains the point $p$. (c) We learn a set of visibility constraints by decomposing the surveillance area into a number of cells $c_1,..,c_{|\mathcal{G}|}$, thus creating a rectangular grid $\mathcal{G}$. For each cell we learn the visible parts of the scene by checking the intersection of the boundary's line segments with the camera-rays for all possible combinations of camera-ray configurations $\mathcal{K}_{\theta_t,x^\text{pos}_t,\xi_t}$.}	
	\label{fig:fig3}
	\vspace{-0mm}
\end{figure*}
\begin{equation}\label{eq:linearSystem}
    \begin{bmatrix}
    \kappa(x)-x_t^\text{pos}(x) &  p_i(x)-p_j(x) \\
    \kappa(y)-x_t^\text{pos}(y) &  p_i(y)-p_j(y)
   \end{bmatrix}
   \begin{bmatrix}
    s\\
    r
   \end{bmatrix} = 
   \begin{bmatrix}
    p_i(x)-\kappa(x)\\
    p_i(y)-\kappa(y)
   \end{bmatrix}.
\end{equation}

\noindent An intersection exists if the pair $(s,r) \in [0,1]$, and the intersection point can be recovered by substituting either $s$ or $r$ into the respective line-segment equations. 

Subsequently, the ray-casting process described above must be performed for each camera-ray $K$, amongst all sets of possible camera-ray configurations $\mathcal{K}_{\theta_t,x^\text{pos}_t,\xi_t}$ and line-segments of the boundary $\Delta \mathcal{C}$, to determine which parts of the scene (i.e., points on the region's boundary), are visible through the agent's camera at each time-step $t \in \mathcal{T}$. This makes the ray-casting computation very computationally expensive. Observe here that the agent's position $x^\text{pos}_t$ is a continuous variable, which depends on the unknown mobility control inputs i.e., Eqn. \eqref{eq:unrolled_kinematics}. More importantly the direct implementation of Eqn. \eqref{eq:1_1}-\eqref{eq:linearSystem} requires the inclusion of non-linear and non-convex constraints in the control problem tackled, which are very hard be handled efficiently. For this reason, in this work an alternative approximate procedure is employed, which allows the ray-casting functionality described above, to be integrated into a mixed integer quadratic program (MIQP) which in turn can be solved to optimality with readily available optimization tools.

In essence the surveillance area $\mathcal{W}$ is first decomposed into a finite number of cells, and then the agent's visible FoV is computed within each cell, for all possible camera-ray configurations. This enables the proposed approach to learn a set of state-dependent visibility constraints which simulate the physical behavior of camera-rays originating within each cell, and which can be embedded into a mixed integer quadratic program. 

Let the rectangular grid $\mathcal{G} = \{c_1,..,c_{|\mathcal{G}|}\}$ to denote the  discretized representation of the surveillance area $\mathcal{W}$, which is composed of cells $c_i, i=1,..,|\mathcal{G}|$ such that $\bigcup_{i=1}^{|\mathcal{G}|} c_i= \mathcal{G}$. To implement the logical conjunction of the visibility constraint in Eqn. \eqref{eq:1_1} we introduce 3 binary variables namely $b^{\mathcal{S}_t}$, $b^{x^\text{pos}_t}$ and $b^{\mathcal{K}}$, which are defined as follows:
\begin{align} 
& b^{\mathcal{S}_t}_{p,m,t} = 1,~ \textit{iff} ~\exists t \in \mathcal{T}, m \in \mathcal{M} : p \in \mathcal{S}_t(m,x^\text{pos}_t) \label{eq:r1},\\
& b^{x^\text{pos}_t}_{c,t} = 1, ~\textit{iff} ~ \exists t \in \mathcal{T} : x^\text{pos}_t \in c \label{eq:r2},\\
& b^{\mathcal{K}}_{c,p} =1,~\textit{iff} ~ \exists K \in \mathcal{K}^c_{\forall \theta,\xi} : (K \otimes \Delta \mathcal{C}) = L_{p}, \label{eq:r3}
\end{align}

\noindent where $\mathcal{M}$ denotes the set of all pairwise combinations $m=(\theta,\xi)$ of $\theta \in \bar{\Theta}$ and $\xi \in \bar{\Xi}$. Hence, the total number of FoV configurations is equal to $|\mathcal{M}| = |\bar{\Theta}| \times |\bar{\Xi}|$ and thus $\mathcal{S}_t(\theta_t,x^\text{pos}_t,\xi_t)$ is abbreviated as $\mathcal{S}_t(m,x^\text{pos}_t), m \in \mathcal{M}$. Above, $c \in \mathcal{G}$ is a rectangular cell, part of the surveillance area $\mathcal{W}$ and $p \in \partial \mathcal{P}$ is a point on the region's boundary $\partial \mathcal{C}$ which is also connected to some line-segment $L_p \in \Delta \mathcal{C}$. 

The binary variable $b^{\mathcal{S}_t}_{p,m,t}$ in Eqn. \eqref{eq:r1}  is activated when the point $p$ is included inside the $m_\text{th} \in \mathcal{M}$ FoV configuration when the agent's position is $x_t^\text{pos}$, at time-step $t$.

Then, the binary variable $b^{x^\text{pos}_t}_{c,t}$ in Eqn. \eqref{eq:r2} shows at which cell $c$ the agent with position $x_t^\text{pos}$ resides at any point in time $t$. Finally the constraint in Eqn. \eqref{eq:r3} indicates with the binary variable $b^{\mathcal{K}}_{c,p}$ if the point $p$ is visible when the agent is inside cell $c$, where the notation $\mathcal{K}^c_{\forall \theta,\xi}$ indicates the sets of camera-ray configurations for all possible combinations of sensor inputs $\theta ~\text{and}~\xi$. 

To be more specific, $b^{\mathcal{K}}_{c,p}$ is learned offline, by pre-computing for each cell $c \in \mathcal{G}$ and for each camera-ray $K \in \mathcal{K}^c_{\forall \theta,\xi}$ the visible part of the boundary $\Delta \mathcal{C}$ via the ray-casting process discussed in the previous paragraph. This is illustrated in Fig. \ref{fig:fig3}(c).

 We should note here that the agent position is sampled uniformly $n_s$ times from within each cell $c$, generating a set of camera-ray configurations $\{\prescript{i}{}{\mathcal{K}^c_{\forall \theta,\xi}}| i=[1,..,n_s]\}$. Therefore for each candidate agent position $x^\text{pos}_i \in i=[1,..,n_s]$ we seek to find the visible points on the boundary. 
 
 Thus, a point $p$ is visible at some time-step $t\in\mathcal{T}$, for some combination of sensor input controls $m \in\mathcal{M}$, and agent position $x^\text{pos}_t$ i.e., $p \in \mathcal{S}^\prime_t(m,x^\text{pos}_t)$ when:

\begin{equation}\label{eq:vcon}
   \exists~ m,t : b^{\mathcal{S}_t}_{p,m,t} \wedge \left[\bigvee_{c=1}^{|\mathcal{G}|} (b^{x^\text{pos}_t}_{c,t} \wedge b^{\mathcal{K}}_{c,p}) \right] = 1.
\end{equation}

\noindent Specifically a point $p$ is visible when both parts of the conjunction in Eqn. \eqref{eq:vcon} are true: a) the point is included inside the agent's sensor FoV which is determined by the agent position $x^\text{pos}$ and sensor control inputs $\theta$ and $\xi$, and encoded by the binary variable $b^{\mathcal{S}_t}_{p,m,t}$ and b)  
the agent with position $x^\text{pos}_t$ resides inside the cell $c\in\mathcal{G}$ (encoded by $b^{x^\text{pos}_t}_{c,t}$) at time $t$, from which originates a camera-ray $K \in \mathcal{K}^c_{\forall \theta,\xi}$ which first intercepts the line $L_p$ which contains point $p$ (i.e., encoded in the learned variable $b^{\mathcal{K}}_{c,p}$). We should note here that the individual terms  of the logical disjunction inside the square brackets are mutually exclusive as the agent cannot occupy two distinct cells at the same time. The ray-casting information encoded in $b^{\mathcal{K}}_{c,p}$ has been learned offline for each cell $c$, and thus during the optimization phase, we need to find the agent's pose which results in a FoV configuration which observes point $p$ (i.e., $b^{\mathcal{S}_t}_{p,m,t}$) and also determine whether the agent resides inside a cell $c \in \mathcal{G}$ (indicated by $b^{x^\text{pos}_t}_{c,t}$) from which point $p$ is visible (as indicated by $b^{\mathcal{K}}_{c,p}$). In the next section, we will show how the above constraints are embedded in the proposed coverage controller.

\begin{algorithm}
\begin{subequations}
\begin{align} 
&\hspace*{-5mm}\textbf{Problem (P2):}~\texttt{Coverage Controller} & \nonumber\\
& \hspace*{-5mm}~~~~\underset{\mathbf{U}_T, \mathbf{\Theta}_T, \mathbf{\Xi}_T}{\arg \min} ~\mathcal{J}_\text{coverage}(\mathbf{X}_T, \mathbf{U}_T, \mathbf{\Theta}_T, \mathbf{\Xi}_T) &\label{eq:objective_P2} \\
&\hspace*{-5mm}\textbf{subject to: $t=[1,..,T]$} ~  &\nonumber\\
&\hspace*{-5mm} x_{t} = \Phi^{t} x_{0} + \sum_{\tau=0}^{t-1} \Phi^{t-\tau-1} \Gamma u_{\tau} & \hspace*{-15mm}\forall t\label{eq:P2_1}\\
&\hspace*{-5mm} \mathcal{V}^\prime_m =  R(\theta)\mathcal{V}_o(\xi) & \hspace*{-15mm} \forall \{(\theta,\xi)\}\in \mathcal{M}  \label{eq:P2_2}\\
&\hspace*{-5mm} \mathcal{V}_{m,t} =  \mathcal{V}^\prime_m + x^\text{pos}_{t} & \hspace*{-15mm} \forall m  \label{eq:P2_3}\\
&\hspace*{-5mm} A^\mathcal{V}_{n,m,t},B^\mathcal{V}_{n,m,t} = \mathcal{L}(\mathcal{V}_{m,t}) & \hspace*{-15mm} \forall m, n=[1,..,3]  \label{eq:P2_4}\\
&\hspace*{-5mm} A^\mathcal{V}_{n,m,t} \times p  ~+  & \label{eq:P2_5}\\
&\hspace*{-5mm}  ~~~~~~b_{n,p,m,t}(M-B^\mathcal{V}_{n,m,t}) \le M & \hspace*{-15mm} \forall n,p,m,t \notag \\ 
&\hspace*{-5mm} 3b^{\mathcal{S}_t}_{p,m,t} - \sum_{n=1}^3 b_{n,p,m,t} \le 0 & \hspace*{-15mm} \forall p,m,t  \label{eq:P2_6}\\
&\hspace*{-5mm} A^\mathcal{\mathcal{G}}_{k,c},B^\mathcal{\mathcal{G}}_{k,c} = \mathcal{L}(\mathcal{G}_{c}) & \hspace*{-15mm} \forall c, k=[1,..,4]  \label{eq:P2_7}\\
&\hspace*{-5mm} A^\mathcal{\mathcal{G}}_{k,c} \times x_t^\text{pos}  ~+  & \label{eq:P2_8}\\
&\hspace*{-5mm}  ~~~~~~\tilde{b}_{k,c,t}(M-B^\mathcal{\mathcal{G}}_{k,c}) \le M & \hspace*{-15mm} \forall k,c,t \notag \\ 
&\hspace*{-5mm} 4b^{x^\text{pos}_t}_{c,t} - \sum_{k=1}^4 \tilde{b}_{k,c,t} \le 0 & \hspace*{-15mm} \forall c,t  \label{eq:P2_9}\\
&\hspace*{-5mm} \sum_{m=1}^{|\mathcal{M}|} \mathcal{F}^\text{sel}_{m,t}=1 & \hspace*{-15mm} \forall t\label{eq:P2_10}\\
&\hspace*{-5mm} b^{\mathcal{S}^\prime_t}_{p,m,t} = ~\mathcal{F}^\text{sel}_{m,t} ~ \wedge & \hspace*{-15mm}  \label{eq:P2_11} \\
&\hspace*{-5mm} ~~~~~~b^{\mathcal{S}_t}_{p,m,t} \wedge \left[\bigvee_{c=1}^{|\mathcal{G}|} (b^{x^\text{pos}_t}_{c,t} \wedge b^{\mathcal{K}}_{c,p}) \right]   & \hspace*{-10mm} \forall p,m,t \notag\\
&\hspace*{-5mm} \sum_{t=1}^T \sum_{m=1}^{|\mathcal{M}|} b^{\mathcal{S}^\prime_t}_{p,m,t} \ge 1 & \forall p\label{eq:P2_12}\\
&\hspace*{-5mm} x_0, x^\text{pos}_{t} \notin \psi,~ \forall \psi \in \Psi & \label{eq:P2_13}\\
&\hspace*{-5mm} x_0, x_{t} \in \mathcal{X}, ~ u_t \in \mathcal{U},~\theta \in \bar{\Theta},~ \xi \in \bar{\Xi}   & \label{eq:P2_13} \notag\\
&\hspace*{-5mm} m \in [1,..,|\mathcal{M}|], p \in [1,..,|\mathcal{P}|], c \in [1,..,|\mathcal{G}|] & \notag\\
&\hspace*{-5mm} \mathcal{F}^\text{sel}_{m,t}, b_{n,p,m,t} \in \{0,1\}  & \notag\\
&\hspace*{-5mm} b^{\mathcal{S}_t}_{p,m,t},\tilde{b}_{k,c,t},b^{x^\text{pos}_t}_{c,t},b^{\mathcal{S}^\prime_t}_{p,m,t},b^{\mathcal{K}}_{c,p} \in \{0,1\}  & \notag
\end{align}
\end{subequations}
\vspace{-8mm}
\end{algorithm}

\subsection{Coverage Controller} \label{ssec:controller}
The complete formulation of the proposed integrated guidance and sensor control coverage planning approach is shown in problem (P2). Specifically, in this section we will show how the high-level problem shown in (P1) is converted into a mixed integer quadratic program (MIQP), which can be solved exactly using readily available optimization tools \cite{Anand2017}. To summarize, our goal in (P2) is to jointly find the mobility and sensor control inputs inside a planning horizon, which optimize a mission-specific objective function i.e., $\mathcal{J}_\text{coverage}$ subject to visibility and coverage constraints. 

\subsubsection{Constraints}
Guidance control is achieved by appropriately selecting the agent's mobility control inputs $\mathbf{U}_T=\{u_t : t \in [0,..,T-1]\}$ governed by its kinematic constraints i.e.  Eqn. \eqref{eq:P2_1}. On the other hand, sensor control is achieved via the constraints in Eqn. \eqref{eq:P2_2}-\eqref{eq:P2_3}. More specifically, in Eqn. \eqref{eq:P2_2} we construct the FoV configurations for all possible pairwise combinations $\{m=(\theta,\xi)\} \in \mathcal{M}$ of the sensor inputs i.e., rotational angle ($\theta$) and zoom-level ($\xi$). Specifically, the continuous variable $\mathcal{V}^\prime_m$ represents a 2 by 3 matrix containing the sensor's FoV vertices for the $m_\text{th}$ FoV configuration. In essence for each zoom-level $\xi \in \bar{\Xi}$ (which determines the FoV parameters $\phi$ and $h$), the FoV is rotated at the origin for each admissible angle $\theta \in \bar{\Theta}$, thus creating a total of $|\mathcal{M}| = |\bar{\Theta}| \times |\bar{\Xi}|$ FoV configurations indexed by $m$ as shown. Subsequently, all the FoV configurations are translated to the agent's position $x^\text{pos}_t$ at time $t$ as shown in Eqn. \eqref{eq:P2_3}. Therefore, the UAV's pose at each time-step inside the planning horizon is completely specified by the constraints in Eqn.\eqref{eq:P2_1}-\eqref{eq:P2_3}.

The constraint in Eqn. \eqref{eq:P2_4}, uses the function $\mathcal{L}(.)$ which takes as input the vertices of the $m_\text{th}$ FoV configuration at some time $t$ i.e., $\mathcal{V}_{m,t}$ and returns a set of linear constraints of the form:
\begin{equation} \label{eq:fov_sys}
     A^\mathcal{V}_{n,m,t} \times x \le B^\mathcal{V}_{n,m,t},
\end{equation}

\noindent where $n=[1,..,3]$ and thus $A^\mathcal{V}_{m,t}$ is a $3$ by 2 matrix, $B^\mathcal{V}_{m,t}$ is a 3 by 1 column vector, and $x$ is column vector representing an arbitrary point in $\mathbb{R}^2$. Given two vertices of the triangular FoV, the function $\mathcal{L}(.)$ works by finding the equation of the line which passes through these two vertices. In total 3 line equations are constructed which fully specify the convex hull of the triangular FoV i.e., a point $x \in \mathbb{R}^2$ belongs to the convex hull of the triangular FoV iff $A^\mathcal{V}_{n,m,t} \times x \le B^\mathcal{V}_{n,m,t}, \forall n=[1,..,3]$. More specifically, $A^\mathcal{V}_{m,t}$ and $B^\mathcal{V}_{m,t}$ contain the coefficients of the lines which form the triangular FoV, where in particular the matrix $A^\mathcal{V}_{n,m,t}$ contains the outward normal to the $n_\text{th}$ line. We can write the line equation which passes from two FoV vertices i.e., $(v^1_x,v^1_y)$ and $(v^2_x,v^2_y)$ as $ax+by=c$, where the coefficients $a = v^1_y - v^2_y$ and $b = v^2_x - v^1_x$ define the normal on the line i.e., $\vec{n} = (a,b)$ and $c = v^2_x v^1_y -v^1_x v^2_y$, thus $A^\mathcal{V}_{1,m,t}=[a, b]$ and $B^\mathcal{V}_{1,m,t}=c$. To summarize, we can determine whether an arbitrary point $x \in \mathbb{R}^2$ is included inside the sensor's FoV by checking if the system of linear inequalities in Eqn. \eqref{eq:fov_sys} is satisfied.

The constraint in Eqn. \eqref{eq:fov_sys} is implemented as shown in (P2) with the constraints shown in Eqn. \eqref{eq:P2_5}-\eqref{eq:P2_6} i.e.,:

\begin{align} 
& A^\mathcal{V}_{n,m,t} \times p + b_{n,p,m,t}(M-B^\mathcal{V}_{n,m,t}) \le M,~ \forall n,p,m,t, \notag\\
& 3b^{\mathcal{S}_t}_{p,m,t} - \sum_{n=1}^3 b_{n,p,m,t} \le 0, ~ \forall p,m,t.\notag
\end{align}

\noindent In essence, these constraints allows us to check whether some point $p \in \mathcal{P}$ is covered by the agent's sensor i.e., $p \in \mathcal{S}_t(m,x^\text{pos}_t)$, when the agent is at position $x^\text{pos}_t$ at time-step $t$, and the sensor's FoV is at the $m_\text{th}$ configuration. As we discussed earlier the matrices $A^\mathcal{V}_{m,t}$ and $B^\mathcal{V}_{m,t}$ contain the coefficients of the sensor's FoV for every possible configuration $m \in \mathcal{M}$ and time-step $t=[1,..,T]$. With this in mind, we use the binary variable $b_{n,p,m,t}$ to decide whether some point $p \in \mathcal{P}$, resides inside the negative half-plane which is created by the $n_\text{th}$ line, of the $m_\text{th}$ FoV configuration at time $t$. When this happens the $b_{n,p,m,t}$ is activated and the inequality in Eqn. \eqref{eq:P2_5} is satisfied i.e., $A^\mathcal{V}_{n,m,t} \times p \le B^\mathcal{V}_{n,m,t}$. On the other hand when $A^\mathcal{V}_{n,m,t} \times p > B^\mathcal{V}_{n,m,t}$, the constraint in Eqn. \eqref{eq:P2_5} is satisfied by setting $b_{n,p,m,t}=0$ and using the large positive constant $M$. Subsequently, the constraint in Eqn. \eqref{eq:P2_6} uses the binary variable $b^{\mathcal{S}_t}_{p,m,t}$ to determine whether the point $p$ resides at time $t$ inside the $m_\text{th}$ configuration of the FoV. Thus $b^{\mathcal{S}_t}_{p,m,t}$ is activated only when $\sum_{n=1}^3 b_{n,p,m,t} = 3$, which signifies that the point $p$ is covered by the sensor's FoV.

Similarly, the next 3 constraints shown in Eqn. \eqref{eq:P2_7} - \eqref{eq:P2_9} (also shown below) use the same principles discussed above, to determine whether the agent with position $x^\text{pos}_t$ resides inside cell $c \in \mathcal{G}$ at time $t$. 

\begin{align} 
&A^\mathcal{\mathcal{G}}_{k,c},B^\mathcal{\mathcal{G}}_{k,c} = \mathcal{L}(\mathcal{G}_{c}), ~\forall k=[1,..,4], c, \notag \\
& A^\mathcal{\mathcal{G}}_{k,c} \times x_t^\text{pos} + \tilde{b}_{k,c,t}(M-B^\mathcal{\mathcal{G}}_{k,c}) \le M , ~ \forall k, c, t,\notag \\
& 4b^{x^\text{pos}_t}_{c,t} - \sum_{k=1}^4 \tilde{b}_{k,c,t} \le 0, ~\forall c, t. \notag
\end{align}

\noindent Specifically, the constraint Eqn. \eqref{eq:P2_7} uses the function $\mathcal{L}(\mathcal{G}_{c})$ on the grid cells and returns in the matrices $A^\mathcal{\mathcal{G}}_{k,c}$ and $B^\mathcal{\mathcal{G}}_{k,c}$, the coefficients of the linear inequalities which define the convex hull of every cell $c \in \mathcal{G}$ in the grid. A point $x\in \mathbb{R}^2$ resides inside a rectangular cell $c$ iff $A^\mathcal{\mathcal{G}}_{c} \times x \le  B^\mathcal{\mathcal{G}}_{c}$, where $A^\mathcal{\mathcal{G}}_{c}$ is a 4 by 2 matrix and $B^\mathcal{\mathcal{G}}_{c}$ is a 4 by 1 column vector. Therefore, the constraint  in Eqn. \eqref{eq:P2_8} uses the binary variable $\tilde{b}_{k,c,t}$ to determine whether the agent's position satisfies the $k_\text{th}$ inequality i.e., $A^\mathcal{\mathcal{G}}_{k,c} \times x_t^\text{pos} \le B^\mathcal{\mathcal{G}}_{k,c}, \forall k=[1,..4]$. Subsequently, the binary variable $b^{x^\text{pos}_t}_{c,t}$ in Eqn. \eqref{eq:P2_9} is activated when $x_t^\text{pos}$ resides inside cell $c$ at time $t$.
Next, we make use of the constraint in Eqn. \eqref{eq:P2_10} i.e.,

\begin{equation}
    \sum_{m=1}^{|\mathcal{M}|} \mathcal{F}^\text{sel}_{m,t}=1, ~\forall t,\notag
\end{equation}

\noindent to account for the fact that at any point in time $t$, only one FoV configuration is active. In other words we would like to prevent the scenario where more than one sets of sensor input controls are applied and executed at some particular time-step $t$. To do this we define the binary variable $\mathcal{F}^\text{sel}$ consisting of $|\mathcal{M}|$ rows and $T$ columns, such that  $\mathcal{F}^\text{sel}_{m,t} \in \{0,1\}, \forall m \in \mathcal{M}, t \in \mathcal{T}$, and we require that at each time-step $t$ only one FoV configuration is active by enforcing the sum of each column to be equal to one, as shown in Eqn. \eqref{eq:P2_10}.

We can now determine whether some point $p$ belongs to the visible FoV as:

\begin{equation}
b^{\mathcal{S}^\prime_t}_{p,m,t} = \mathcal{F}^\text{sel}_{m,t} \wedge b^{\mathcal{S}_t}_{p,m,t} \wedge \left[\bigvee_{c=1}^{|\mathcal{G}|} (b^{x^\text{pos}_t}_{c,t} \wedge b^{\mathcal{K}}_{c,p}) \right],~ \forall p,m,t, \notag
\end{equation}

\noindent where the binary variable $b^{\mathcal{S}^\prime_t}_{p,m,t}$ is activated when the point $p \in \mathcal{P}$ is visible at time $t$, and specifically resides inside the $m_\text{th}$ FoV configuration i.e., $p \in \mathcal{S}_t^\prime(m, x^\text{pos}_t)$. As it is shown, in the conjunction above, the binary variable $b^{\mathcal{S}_t}_{p,m,t}$ captures the sensor's pose, which is determined by the agent's position $x_t^\text{pos}$ at time $t$ and sensor orientation given by the $m_\text{th}$ FoV configuration, and determines whether the point $p$ resides inside the sensor's FoV i.e., constraints in Eqn. \eqref{eq:P2_5}-\eqref{eq:P2_6}. Then the binary variable $b^{x^\text{pos}_t}_{c,t}$ checks if the agent at time $t$ resides inside a particular cell $c$ i.e., Eqn. \eqref{eq:P2_7} - \eqref{eq:P2_9} and together with the learned variable $b^{\mathcal{K}}_{c,p}$ (as discussed in Sec. \ref{ssec:vis_con}), determine whether the point $p$ is visible given that the agent is at cell $c$. To summarize, a point $p$ belongs to the visible FoV when there exists some cell $c$ from which the point $p$ is visible and at time-step $t$ the agent's position $x^\text{pos}_t$ resides inside that cell $c$ and exists a FoV configuration $m$ such that point $p$ is covered by the agent's sensor at time $t$. Lastly, in order to make sure that only one FoV configuration is active at each time-step (i.e., one set of sensor controls is applied), we use the FoV selector $\mathcal{F}^\text{sel}_{m,t}$ as shown in Eqn. \eqref{eq:P2_10}.

Finally, we can ensure that during the planning horizon, every point $p \in \mathcal{P}$ will be covered at least once by the agent's sensor via the constraint in Eqn. \eqref{eq:P2_12}:
\begin{equation}
    \sum_{t=1}^T \sum_{m=1}^{|\mathcal{M}|} b^{\mathcal{S}^\prime_t}_{p,m,t} \ge 1, ~\forall p, \notag
\end{equation}

\noindent where we require that there exists at least one FoV configuration (i.e., one set of sensor input controls) $m \in \mathcal{M}$ at some time-step $t \in \mathcal{T}$, such that the binary variable $b^{\mathcal{S}^\prime_t}_{p,m,t}$ is activated i.e., $b^{\mathcal{S}^\prime_t}_{p,m,t}=1$ for every point $p \in \mathcal{P}$. Therefore, determining the agent's mobility and sensor control inputs such that all points are included inside the agent's visible FoV at some time-step during the planning horizon.

The last constraint shown in Eqn. \eqref{eq:P2_13}, implements the obstacle avoidance constraints as discussed in Sec. \ref{ssec:preliminaries}. To summarize, we have derived a set of constraints i.e., Eqn.\eqref{eq:P2_1}-\eqref{eq:P2_13} which jointly account for the agent's kinematic and sensing model, integrate visibility constraints into the coverage planning problem and guarantee full coverage inside the planning horizon, given that a feasible solution exists. Finally, we should note that the formulation of the coverage controller shown in problem (P2) can be easily extended to tackle the problem in 3D environments, however this has been left for future works. Next, we discuss in detail the design of the objective function, which can be used in order to optimize a set of mission-related performance criteria.

\subsubsection{Objectives}
The problem of integrated UAV guidance and sensor control which is studied in this work is a core component for many applications and tasks including surveillance, emergency response, and search-and-rescue missions. Take for instance a UAV-based search-and-rescue mission where the objective is to search an area of interest and locate as quickly as possible people in need. In such scenarios, the mission's completion time is of the highest importance for saving lives. In other cases the UAV's  efficient battery utilization is imperative for the success of the mission. Motivated by the objectives and requirements discussed above we design a multi-objective cost function $\mathcal{J}_\text{coverage}$ to allow for the characterization of several mission-related optimality criteria, and trade-offs amongst these. In this work, $\mathcal{J}_\text{coverage}$ is composed of a set of sub-objectives, which sometimes might be competing. More specifically, we define the overall coverage objective $\mathcal{J}_\text{coverage}$ as:
\begin{equation}
    \mathcal{J}_\text{coverage} = \left( w_1J_1 + w_2J_2+\ldots+w_nJ_n \right),
\end{equation}

\noindent where $J_i$ represents the $i_\text{th}$ sub-objective and $w_i$ is the tuning weight associated with the $i_\text{th}$ sub-objective. Therefore, the weights are used to emphasize or deemphasize the importance of each sub-objective according to the mission goals. Next we design several possible sub-objectives which can be used to drive an efficient coverage planning mission.

\begin{figure*}
	\centering
	\includegraphics[width=\textwidth]{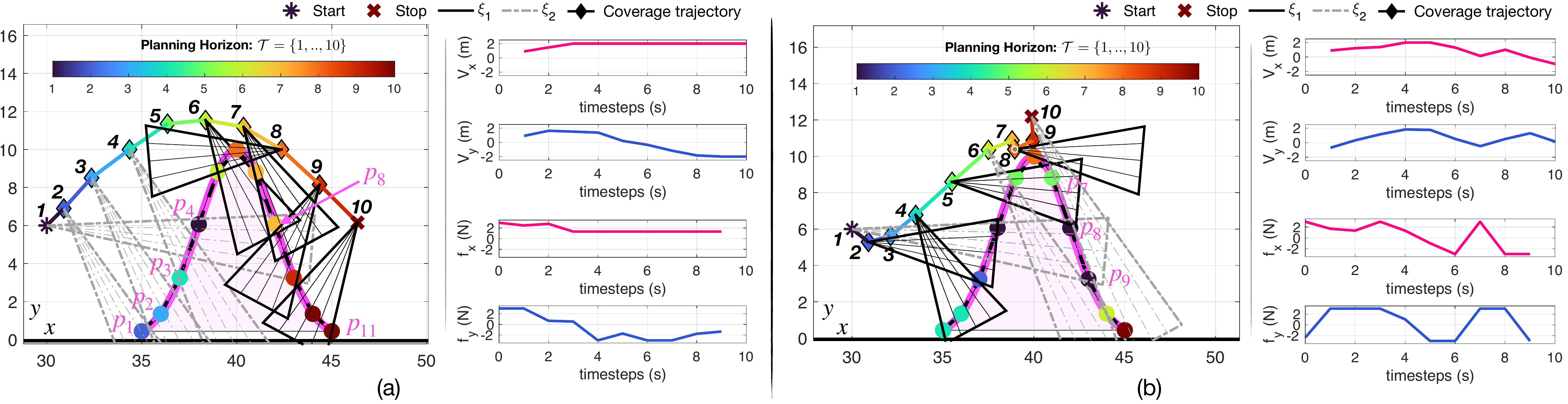}
	\caption{The figure illustrates the impact of the visibility constraints on the generated coverage trajectories. (a) Visibility constraints are enabled, (b) Visibility constraints disabled. The visibility constraints simulate the physical behavior of camera-rays, therefore allowing the agent to distinguish between visible and occluded points.}	
	\label{fig:res1}
	\vspace{-0mm}
\end{figure*}

\noindent \textbf{Mission completion time ($J_1$):} As discussed earlier one of the most important  objectives in a coverage planning scenario is the minimization of the mission's completion time. In other words, we are interested in finding the optimal UAV control inputs (i.e., mobility and sensor controls), which when executed allow the agent to cover all points of interest $p \in \mathcal{P}$ as quickly as possible, thus minimizing the time needed to conduct a full coverage of the region of interest. This can be defined as follows:

\begin{equation}\label{eq:j1}
    J_1 = \sum_{p=1}^{|\mathcal{P}|}\sum_{m=1}^{|\mathcal{M}|}\sum_{t=1}^{T} \left( b^{\mathcal{S}^\prime_t}_{p,m,t} \times \frac{t}{T} \right).
\end{equation}

\noindent In essence by minimizing $J_1$, we are minimizing the product of the binary variable $b^{\mathcal{S}^\prime_t}_{p,m,t}$ with the factor $(t/T)$, over the planning horizon of length $T$, for all points $p \in \mathcal{P}$ and FoV configurations $m \in \mathcal{M}$. Effectively, $(t/T)$ in Eqn. \eqref{eq:j1} acts as a weight to $b^{\mathcal{S}^\prime_t}_{p,m,t}$ which increases over time. This drives the optimizer to find the optimal mobility and sensor control inputs which allow the agent to cover all points $p \in \mathcal{P}$ as quickly as possible or equivalently $b^{\mathcal{S}^\prime_t}_{p,m,t}$ is activated for each point at the earliest possible time-step. Finally, we should note here that the agent's control inputs are directly linked with $b^{\mathcal{S}^\prime_t}_{p,m,t}$, since the agent's pose is jointly determined by its mobility and sensor controls i.e., Eqn. \eqref{eq:P2_2}-\eqref{eq:P2_3} and also $b^{\mathcal{S}^\prime_t}_{p,m,t}$ is only activated when point $p$ is visible.

\noindent \textbf{Energy Efficiency ($J_2$):} Energy-aware operation is another essential objective for various applications. In essence we are interested in prolonging the UAV's operation time (i.e., minimizing the battery drain), by optimizing the UAV's mobility control inputs (i.e., the amount of force applied), thus generating energy-efficient coverage trajectories. Although, the proposed coverage planning formulation, does not directly uses a battery model for the UAV, it is assumed that the UAV mobility control inputs are directly linked with the battery usage. Therefore, energy efficient coverage planning can be achieved by appropriately selecting the UAV's mobility control inputs. Specifically, it is assumed that the generation of smooth UAV trajectories with reduced abrupt changes in direction and speed can lead to improved battery usage, thus we define the energy-aware coverage planning sub-objective as:

\begin{equation}\label{eq:j2}
    J_2 = \sum_{t=1}^{T-1} ||u_t - u_{t-1}||^2_2 + \sum_{t=0}^{T-1} |u_t|,
\end{equation}

\noindent where we minimize a) the sum of deviations between consecutive control inputs and b) the cumulative magnitude of the absolute value of individual controls, thus leading to energy optimized coverage planning.

\noindent \textbf{Sensor Control Effort ($J_3$):} The last objective aims to minimize the sensor deterioration due to excessive and/or improper usage i.e., by reducing the utilization of the gimbaled sensor during the mission. This allows us to maintain the sensor's healthy status and prolong its lifespan. For this reason, we define as sensor control effort the deviation between successive FoV configurations  and thus $J_3$ is defined by:

\begin{equation}\label{eq:j3}
    J_3 = \sum_{t=1}^{T-1} \sum_{m=1}^{|\mathcal{M}|}||\mathcal{F}^\text{sel}_{m,t+1} - \mathcal{F}^\text{sel}_{m,t}||^2_2,
\end{equation}

\noindent which favors the generation of coverage trajectories which achieve full coverage with minimum gimbal utilization.

To summarize, in this section we have described a set (not exhaustive) of sub-objectives, which can be used to compose the overall multi-objective cost function $\mathcal{J}_\text{coverage}$ for the coverage path planning problem we examine in this work. These sub-objectives can be prioritized depending on the problem requirements, while new ones can be added according to the mission specifications.
We should mention here that the objective $J_3$ described above can also be incorporated into the objective $J_2$ i.e., energy efficiency, to account for the overall energy expenditure (i.e., energy expenditure from motion control and from gimbal control) of the system.

%% file: evaluation.tex
\section{Evaluation} \label{sec:Evaluation}
\subsection{Simulation Setup}

In order to evaluate the proposed integrated guidance and gimbal control coverage approach we have conducted a thorough simulation analysis. More specifically, the evaluation is divided into three parts. In the first part we investigate the effect of the visibility constraints on the coverage planning behavior. In the second part of the evaluation we showcase the proposed approach for various mission related optimality criteria, and finally in the third part, we analyze the generated coverage trajectories for various parameters of the inputs. 

The simulation setup used for the evaluation of the proposed approach is as follows: The agent kinematics are expressed by Eqn. \eqref{eq:agent_dynamics} with $\delta t=1$s, agent mass $m=3.35$kg and drag coefficient $\eta=0.2$. For demonstration purposes, the control input (i.e., input force) $u_t=[f_t(x),f_t(y)]$ is bounded in each dimension according to $|f_t(x|y)| \le 3$N, and the agent velocity is bounded according to $|\nu_t(x|y)|\le 2$m/s. The agent's FoV angle is set to $\phi=30$deg, and the sensing range $h=7$m. The camera zoom-levels are set to $\bar{\Xi}=\{1,2\}$, thus the camera characteristics for zoom-level $\xi=1$ and $\xi=2$ are given by $(\phi=30,h=7)$ and $(\phi=15,h=14)$ respectively. In total we consider 4 camera rotation angles i.e., $\bar{\Theta}=\{-85, -28, 28, 85\}$ which are used to rotate the camera FoV according to Eqn. \eqref{eq:rotation_mat}, leading to a total of 8 possible FoV configurations i.e., $|\mathcal{M}|=8$. We have used the ray-tracing procedure with $|\mathcal{K}|=5$ camera-rays in a surveillance region $\mathcal{W}$ that has a total area of $60 \times 20 \text{m}^2$. The region/object of interest $\mathcal{C}$ is represented by a bell-shaped curve (as illustrated in Fig. \ref{fig:fig2}), given by $f(x) = a\times\text{exp}\left(\frac{-(x-b)^2}{2c^2} \right)$ with $a, b$ and $c$ set to 10, 40 and 2 respectively. The region of interest is assumed to be non traversable, and thus we are interested in generating coverage trajectories to cover a total of 11 points $\mathcal{P}=\{p_1,..,p_{11}\}$ sampled from the region's boundary $\partial \mathcal{C}$. Finally, we should mention that the visibility constraints have been learned on a discretized representation $\mathcal{G}$ of the surveillance area, where $\mathcal{G}$ contains 16 square cells (as illustrated in Fig. \ref{fig:fig3}) of size $10\text{m} \times 5 \text{m}$. To summarize, the visibility constraints have been learned according to Sec. \ref{ssec:vis_con}, with $|\mathcal{M}|=8, |\mathcal{P}|=11, |\mathcal{G}|=16, n_s=15$ and $|\Delta\mathcal{C}|=11$. The results have been obtained with Gurobi v9 solver, running on a 2.5GHz laptop computer.

\begin{figure*}
	\centering
	\includegraphics[width=\textwidth]{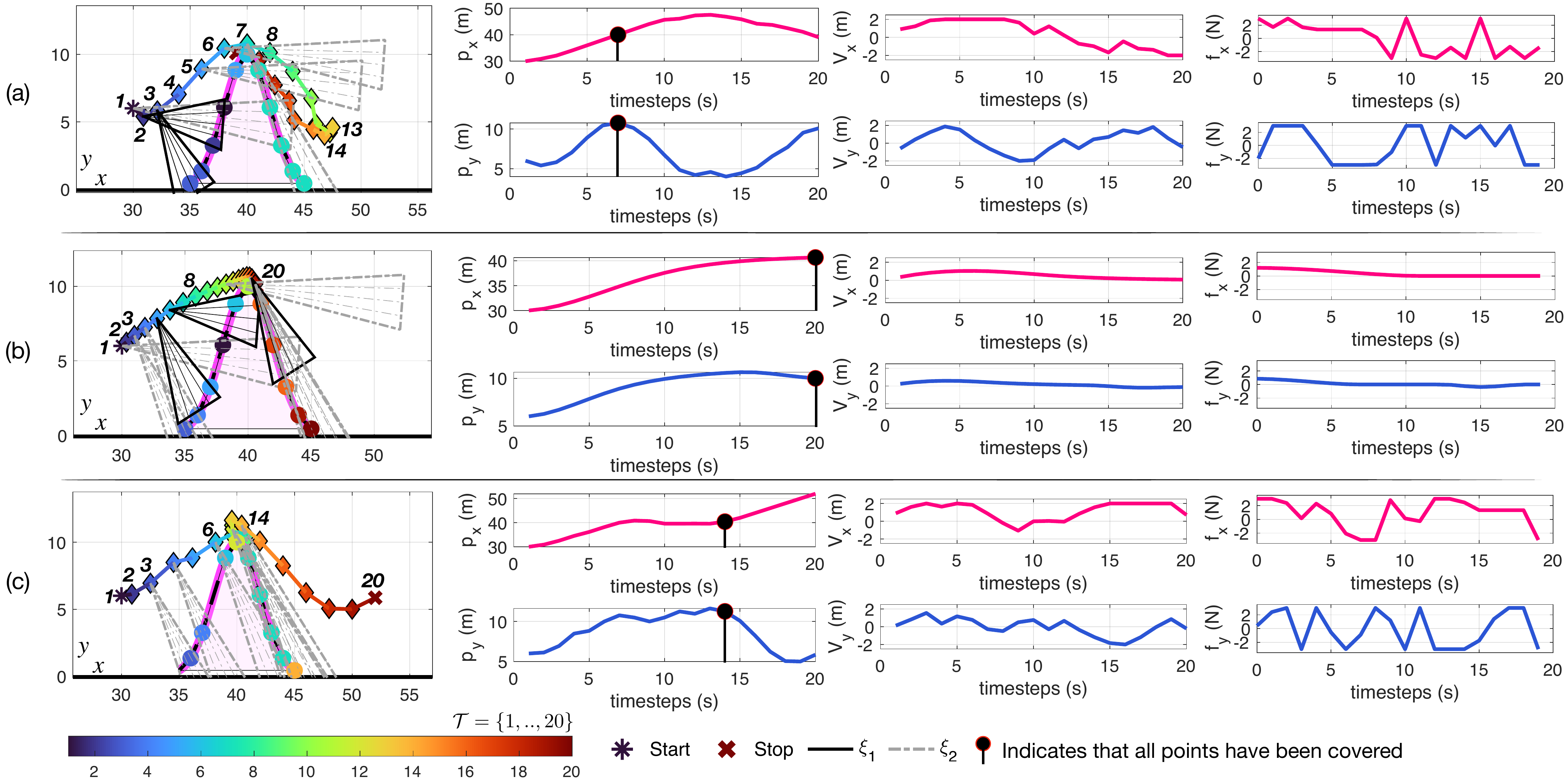}
	\caption{The figure illustrates the generated coverage planning trajectories over a planning horizon of length 20 time-steps, for 3 different sub-objectives. (a) Mission completion time ($J_1$), (b) Energy efficiency ($J_2$), and (c) Sensor control effort ($J_3$).}	
	\label{fig:res2}
	\vspace{-0mm}
\end{figure*}

\subsection{Results}
\subsubsection{\textbf{Visibility Constraints}} With the first experiment, shown in Fig. \ref{fig:res1}, we aim to investigate the impact of the visibility constraints on the trajectory generation process, and gain insights on the coverage planning behavior of the proposed approach. Specifically, Fig. \ref{fig:res1}(a) shows the coverage trajectory, agent velocity, and mobility control inputs, within a planning horizon of $T=10$ time-steps when the visibility constraints are enabled, whereas Fig. \ref{fig:res1}(b) shows the exact same scenario with the visibility constraints disabled. The region of interest is shaded in pink (i.e., the bell-shaped curve), the agent's trajectory is marked with $-\diamondsuit-$, the agent's start and stop positions are marked with $\star$ and $\times$ respectively, and the points on the boundary to be covered are marked with $\bullet$. The figure also shows the FoV configuration at each time-step, indicated by the isosceles triangles, where the black solid lines and the gray dashed lines correspond to the first ($\xi_1$) and second ($\xi_2$) zoom-levels respectively. Finally, the agent's trajectory and the points to be covered are color-coded according to the time-step which are observed, as shown in the figure legend. Therefore, according to Fig. \ref{fig:res1}(a), point $p_1$ (colored dark blue) is covered at time-step $t=2$, point $p_2$ (colored light blue) is covered at $t=3$, point $p_3$ (colored light green) is included inside the agent's FoV at $t=4$, point $p_4$ (colored black) is the first point to be viewed by the agent at time-step $t=1$, and so on and so forth. We should mention that for this experiment we have set $\mathcal{J}_\text{coverage} = 1$ (i.e., we are minimizing a constant), and thus the solely goal of the optimization in this experiment is to satisfy the coverage constraints (i.e., cover all points). 

As we can observe from Fig. \ref{fig:res1}(a), the agent starts from the left side of the bell-shaped curve, and appropriately selects its mobility and sensor control inputs which achieve full coverage. More importantly, we can observe that although the FoV can extend all the way through the object of interest (e.g., at $t=1$ points $p_4$ and $p_8$ are inside the FoV), the use of visibility constraints, which simulate the physical behavior of camera-rays, allow the identification of occlusions (e.g., at $t=1$ point $p_8$ is occluded, and becomes visible at $t=7$ as shown). Therefore, the agent can identify at each time-step which points are visible through its camera and plan its coverage trajectory as needed. For this reason, in this test the agent goes over the bell-shaped curve, towards the other side of the curve, in order to cover the occluded points i.e., at $t=7$ points $p_7$ and $p_8$ are covered, at $t=8$ point $p_6$ is covered, at $t=9$ the point $p_9$ is covered and the remaining points ($p_{10}$ and $p_{11}$) are covered at time-step $t_{10}$. In addition, it is shown that the obstacle avoidance constraints restrict the agent from passing through the object of interest. On the other hand, observe from Fig. \ref{fig:res1}(b) that when the visibility constraints are disabled, the agent cannot distinguish between visible and occluded parts of the scene i.e., at $t=1$ the points $p_8$ and $p_9$ (colored black) are occluded but observed, similarly point $p_7$ (colored green) is occluded at $t=5$ but it is observed as shown in the figure. This is because the sensor's visible FoV is not modeled adequately without the use of the visibility constraints, and as a result the generated trajectory does not resembles a realistic coverage path.

\subsubsection{\textbf{Coverage Objectives}}
The purpose of the next experiment is to investigate in more detail different coverage planning strategies by optimizing the sub-objectives discussed in Sec. \ref{ssec:controller}. More specifically, we will show how the coverage plan changes when optimizing the mission completion time ($J_1$), the energy efficiency ($J_2$), the sensor control effort ($J_3$), and a weighted combination of those. Figure \ref{fig:res2} shows the coverage planning trajectories along with the agent position, velocity, and input force over time for the same scenario, when optimizing the individual sub-objectives $J_1, J_2,$ and $J_3$, within a planning horizon of length 20 time-steps. As it can be seen from Fig. \ref{fig:res2}(a), when optimizing the mission completion time ($J_1$), the set of 11 points $\mathcal{P}$ is fully covered at time-step 7. Note that the agent's trajectory and the points to be covered are color-coded according to the time-step at which the coverage occurs. In this sense the last point in Fig. \ref{fig:res2}(a) which is color-coded light green is covered at time-step 7. The time-step at which all points are covered is also shown in the agent position plot, and marked with a black circle. We should point out here that for visual clarity the graphs show the FoV configurations only for the time-instances for which a point is included inside the sensor's FoV. Next, Fig. \ref{fig:res2}(b) shows the coverage trajectory for the sub-objective which minimizes the agent's energy expenditure. As it is shown, in this scenario the applied input force which is used for guidance is driven to zero over time. In addition we can observe that the agent moves in small increments (i.e., consecutive positions are close to each other) as opposed to the previous scenario, also indicated by the velocity plot. Also, observe how the agent utilizes its sensor to achieve full coverage (which occurs at time-step 20), while optimizing for energy efficiency. In this scenario, 6 out of the 8 camera FoV configurations are used over the planning horizon, until full coverage is achieved. Finally, Fig. \ref{fig:res2}(c), shows that the minimization of the sensor control effort ($J_3$), forces the agent to complete the mission by utilizing just 1 out of the 8 possible FoV configurations. Essentially, in this scenario the camera remains fixed as shown in the figure. In this scenario full coverage was achieved at time-step 14 as shown in the graph. Also observe that in this scenario the mobility controls fluctuate significantly as opposed to the previous scenario.

\begin{figure}
	\centering
	\includegraphics[width=\columnwidth]{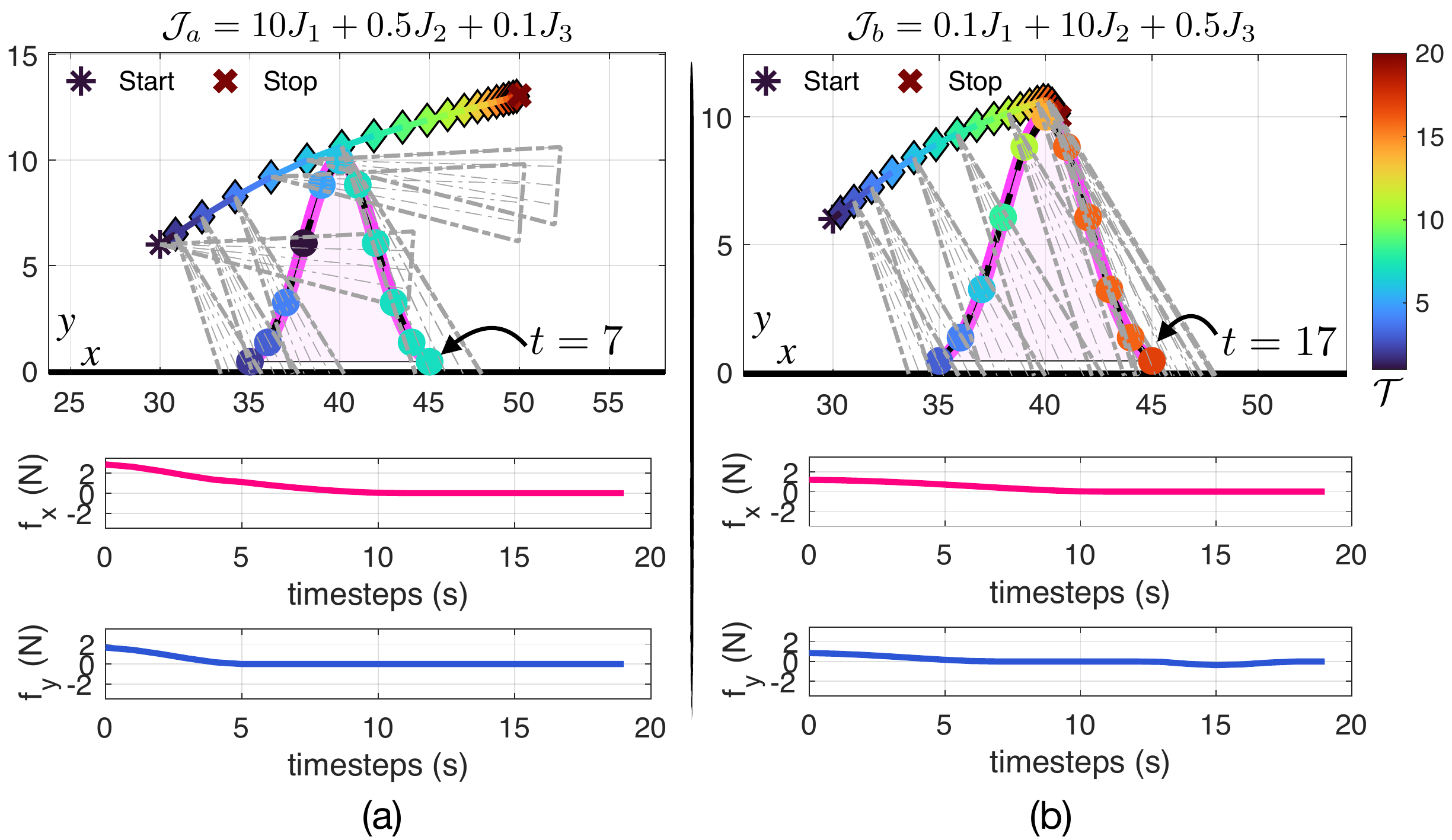}
	\caption{The figure illustrates the coverage planning trajectories over a planning horizon of 20 time-steps, for two different multi-objective cost functions. (a) $\mathcal{J}_a = 10J_1+0.5J_2+0.1J_3$, and (b) $\mathcal{J}_b = 0.1J_1+10J_2+0.5J_3$.}	
	\label{fig:res3}
	\vspace{-0mm}
\end{figure}

These three different sub-objectives can be incorporated into a multi-objective cost function, as discussed in Sec. \ref{ssec:controller}, and by adjusting the emphasis given to each sub-objective the desired coverage behavior can be obtained as shown in the next experiment i.e., Fig. \ref{fig:res3}. More specifically, we have run the coverage planning with two different multi-objective cost functions i.e., $\mathcal{J}_a = 10J_1+0.5J_2+0.1J_3$ and $\mathcal{J}_b = 0.1J_1+10J_2+0.5J_3$ shown in Fig. \ref{fig:res3}(a) and Fig. \ref{fig:res3}(b) respectively. In both scenarios we aim to optimize for energy efficiency by including, and appropriately weighting sub-objective $J_2$ in the multi-objective cost function. As it is illustrated the applied force is minimized and driven to zero for both scenarios. However, due to the higher emphasis given to $J_2$ in the second scenario, $\mathcal{J}_b$ optimizes energy savings more aggressively also indicated by the generated coverage trajectory. Observe that in objective $\mathcal{J}_a$ greater emphasis is given to mission completion time (i.e., $J_1$), which results in faster coverage (at time-step 7). On the other hand in $\mathcal{J}_b$ full coverage is achieved at time-step 17. Finally, we observe that by weighting more on the sub-objective $J_3$ in $\mathcal{J}_b$, allows the agent minimize the gimbal rotations by maintaining the camera fixed and at the same time optimize for energy efficiency as shown in Fig. \ref{fig:res3}(b).

\begin{figure}
	\centering
	\includegraphics[width=\columnwidth]{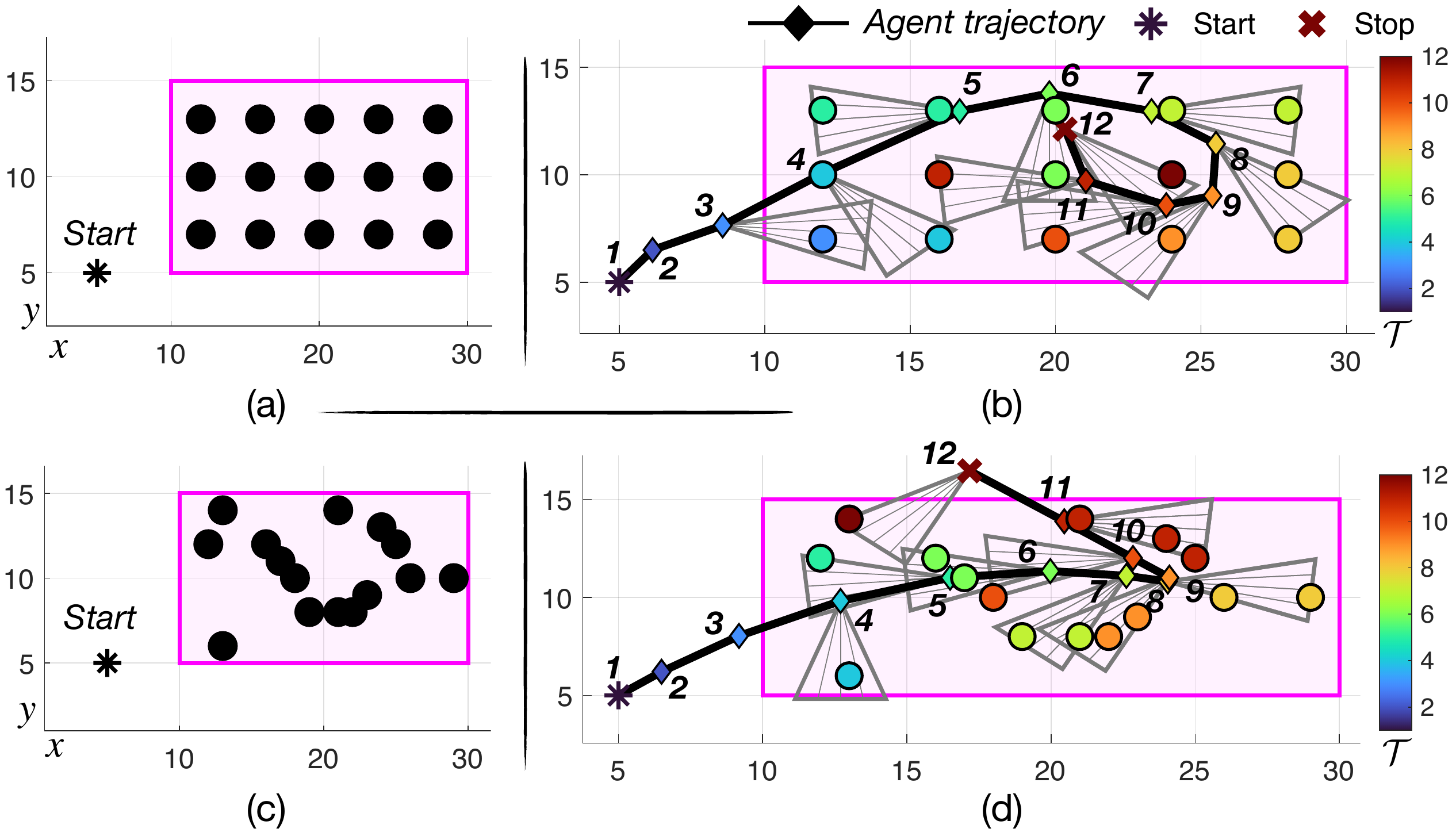}
	\caption{The figure illustrates the coverage plan for two different scenarios which exhibit a traversable area of interest, indicated by the pink shaded rectangle. (a)(b) The agent's coverage plan when the area of interest is approximated by $|\mathcal{P}|=15$ equally spaced points shown as $\bullet$, (c)(d) The agent's trajectory used to cover $|\mathcal{P}|=15$ points uniformly sampled from the area of interest. The numbers on the trajectory indicate time-steps.}	
	\label{fig:res5}
	\vspace{-0mm}
\end{figure}

Our next experiment shows the behavior of the proposed coverage planning approach for a traversable region of interest, indicated by the pink shaded rectangle shown in Fig. \ref{fig:res5}. In this scenario, without loss of generality, we assume an obstacle-free region of interest. More specifically, in Fig. \ref{fig:res5}(a) the region of interest is approximated with a total of $|\mathcal{P}|=15$ equally spaced points (shown as $\bullet$), which need to be covered by the agent, initially located at $(x,y)=(5,5)$ (shown as $\star$). Figure \ref{fig:res5}(b) shows the agent's coverage trajectory when optimizing the coverage objective $\mathcal{J} = 0.1J_1 + J_2$ inside a planning horizon of length $T=12$ time-steps. In this scenario, the agent's FoV angle $\phi$ and sensing range $h$ are set to $35$deg, and $5$m respectively, and the camera can be rotated in 5 ways i.e., $\bar{\Theta}=\{-85, -42.5, 0, 42.5, 85\}$. In order to make the illustrations easier to read, in this scenario we do not make use of the zoom functionality i.e., $\bar{\Xi}=\{1\}$ which leads to a total of 5 possible FoV configurations i.e., $|\mathcal{M}|=5$. As shown in Fig. \ref{fig:res5}(b) the agent's mobility control inputs and camera rotations are appropriately selected and optimized according to the coverage objective $\mathcal{J}$ to achieve full coverage of the region of interest, i.e., as shown in the figure the points are colored-coded according to the time-step they are observed by the agent. Next, Fig. \ref{fig:res5}(c)(d) shows the same setup for 15 points which are sampled uniformly from the region of interest. As shown in Fig. \ref{fig:res5}(d) the generated coverage plan enables the agent to cover all 15 points over the planning horizon.

\begin{figure}
	\centering
	\includegraphics[width=\columnwidth]{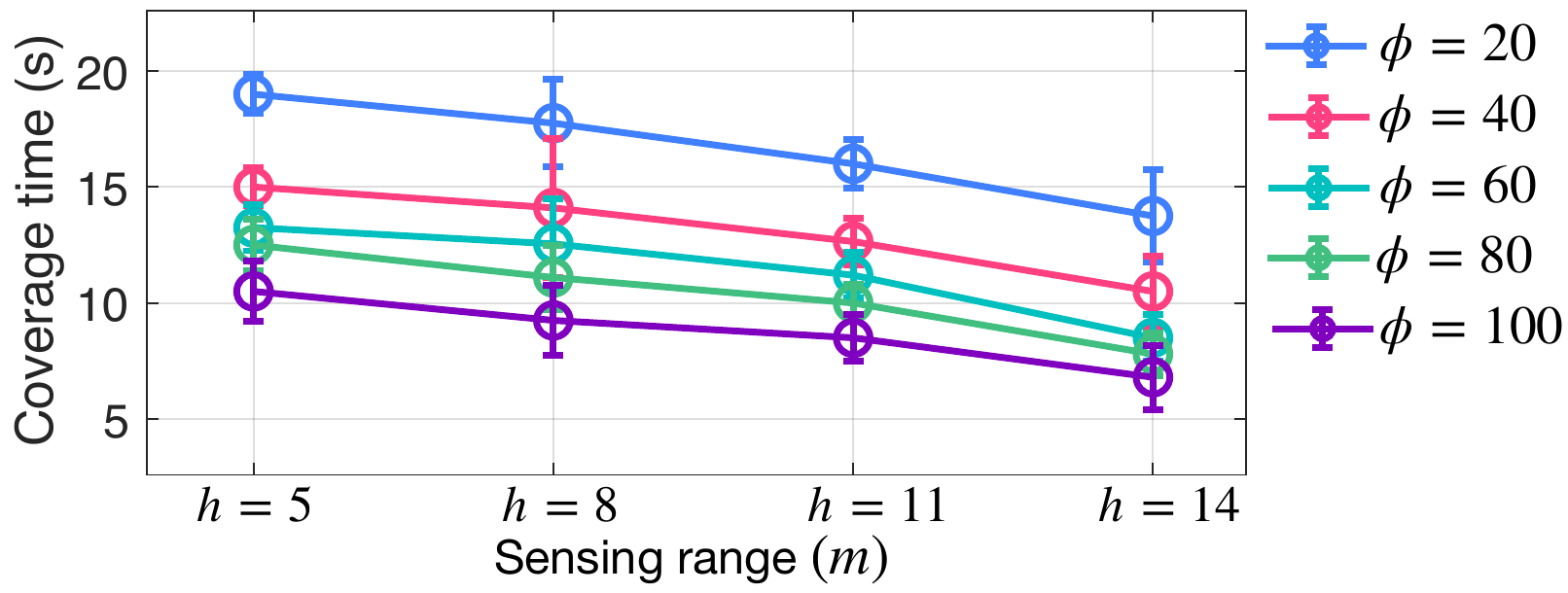}
	\caption{The figure shows the mission completion time for various parameter configurations of the FoV, i.e., $\phi \in \{20, 40, 60, 80, 100\}$ and $h \in \{5, 8, 11, 14\}$.}	
	\label{fig:res4}
	\vspace{-0mm}
\end{figure}

 Our last experiment investigates how various configurations of the FoV parameters $\phi$ and $h$ (i.e., opening angle and sensing range) affect the coverage performance and more specifically the mission completion time. For this experiment, we use the simulation setup discussed in the beginning of this section, with $\bar{\Xi}=\{1\}$ and optimizing $J_1$. We perform 50 Monte Carlo trials, where we sample the agent position randomly within a disk of radius 5m centered at $(x,y)=(30,6)$ for each combination of the parameters $(\phi,h) \in \Phi \times H$, where $\Phi \in \{20, 40, 60, 80, 100\}$deg and $H = \{5, 8, 11, 14\}$m. In particular, Fig. \ref{fig:res4} shows the average coverage completion time for all configurations of the parameters. As we can observe the time needed for full coverage drops from 20sec to below 15sec when we increase the sensing range from $h=5$m to $h=14$m for the scenario where the angle opening is set to $\phi=20$ deg. Similarly, for fixed sensing range set at $h=5$m, the coverage time reduces approximately by 50$\%$ as the angle opening increases from 20deg to 100deg. Overall, as we can observe the from Fig. \ref{fig:res4} the mission completion time improves as the FoV increases both in terms of $\phi$ and $h$. Finally, observe that the extremities of the bell shaped object of interest are 10m tall (at the very top) and 10m wide (at the base), which means that without the use of the proposed visibility constraints and with a camera configuration of $\phi=100$deg and $h=14$m, the UAV agent could have observed all points and finish the mission in a couple of seconds, as the entire object of interest would have been included inside it's sensor's footprint. However, this erroneous behavior is prevented in this work by the integration of ray-casting into the proposed coverage controller.

  %
%

%% file: conclusion.tex
\section{Conclusion} \label{sec:conclusion}

In this work we have proposed an integrated guidance and gimbal control approach for coverage path planning. In the proposed approach the UAV's mobility and sensor control inputs are jointly optimized to achieve full coverage of a given region of interest, according to a specified set of optimality criteria including mission completion time, energy efficiency and sensor control effort. We have devised a set of visibility constraints in order to integrate ray-casting to the proposed coverage controller, thus allowing the generation of optimized coverage trajectories according to the sensor's visible field-of-view. Finally, we have demonstrated how the constrained optimal control problem tackled in this work can be formulated as a mixed integer quadratic program (MIQP), and solved using off-the-shelf tools. Extensive numerical experiments have demonstrated the effectiveness of the proposed approach. Future directions include the extension of the proposed approach in 3D environments, the evaluation of the proposed approach in real-world settings, and extensions to multiple agents.

%% file: ack.tex
\section*{Acknowledgments}

This work is supported by the European Union's Horizon 2020 research and innovation programme under grant agreement No 739551 (KIOS CoE), from the Republic of Cyprus through the Directorate General for European Programmes, Coordination and Development, and from the Cyprus Research and Innovation Foundation under grant agreement EXCELLENCE/0421/0586 (GLIMPSE).

%% file: biography.tex
\begin{IEEEbiography}[{\includegraphics[width=1in,height=1.25in,clip,keepaspectratio]{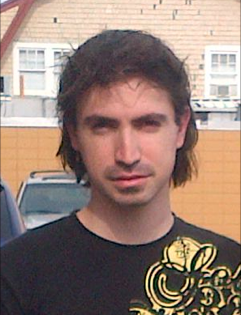}}]%
{Savvas Papaioannou} received the B.S. degree in Electronic and Computer Engineering from the Technical University of Crete, Chania, Greece in 2011, the M.S. degree in Electrical Engineering from Yale University, New Haven, CT, USA, in 2013, and the Ph.D. degree in Computer Science from the University of Oxford, Oxford, U.K. in 2017. He is currently a Research Associate with the KIOS Research and Innovation Center of Excellence, University of Cyprus, Nicosia, Cyprus. His research interests include multi-agent and autonomous systems, state estimation and control, multi-target tracking, trajectory planning, and intelligent unmanned aerial vehicle (UAV) systems and applications. Dr. Papaioannou is a reviewer for various conferences and journals of the IEEE and ACM Associations, and he has served in the organizing committees of various international conferences.
\end{IEEEbiography}

\begin{IEEEbiography}[{\includegraphics[width=1in,height=1.25in,clip,keepaspectratio]{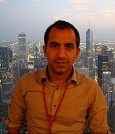}}]%
{Panayiotis Kolios} is currently a Research Assistant Professor at the KIOS Research and Innovation Centre of Excellence of the University of Cyprus. He received his BEng and PhD degree in Telecommunications Engineering from King's College London in 2008 and 2011, respectively. Before joining the KIOS CoE, he worked at the Department of Communications and Internet Studies at the Cyprus University of Technology and the Department of Computer Science of the University of Cyprus (UCY). His work focuses on both basic and applied research on networked intelligent systems. Some examples of systems that fall into the latter category include intelligent transportation systems, autonomous unmanned aerial systems and the plethora of cyber-physical systems that arise within the Internet of Things. Particular emphasis is given to emergency management in which natural disasters, technological faults and man-made attacks could cause disruptions that need to be effectively handled. Tools used include graph theoretic approaches, algorithmic development, mathematical and dynamic programming, as well as combinatorial optimization.
\end{IEEEbiography}

\begin{IEEEbiography}[{\includegraphics[width=1in,height=1.25in,clip,keepaspectratio]{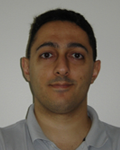}}]%
{Theocharis Theocharides} is an Associate Professor in the Department of Electrical and Computer Engineering, at the University of Cyprus and a faculty member of the KIOS Research and Innovation Center of Excellence where he serves as the Research Director. Theocharis received his Ph.D. in Computer Engineering from Penn State University, working in the areas of low-power computer architectures and reliable system design, where he was honored with the Robert M. Owens Memorial Scholarship, in May 2005. He has been with the Electrical and Computer Engineering department at the University of Cyprus since 2006, where he directs the Embedded and Application-Specific Systems-on-Chip Laboratory. His research focuses on the design, development, implementation, and deployment of low-power and reliable on-chip application-specific architectures, low-power VLSI design, real-time embedded systems design and exploration of energy-reliability trade-offs for Systems on Chip and Embedded Systems. His focus lies on acceleration of computer vision and artificial intelligence algorithms in hardware, geared towards edge computing, and in utilizing reconfigurable hardware towards self-aware, evolvable edge computing systems. His research has been funded by several National and European agencies and the industry, and he is currently involved in over ten funded ongoing research projects. He serves on several organizing and technical program committees of various conferences (currently serving as the Application Track Chair for the DATE Conference), is a Senior Member of the IEEE and a member of the ACM. He  is currently an Associate Editor for the ACM Transactions on Emerging Technologies in Computer Systems, IEEE Consumer Electronics magazine, IET's Computers and Digital Techniques, the ETRI journal and Springer Nature Computer Science. He also serves on the Editorial Board of IEEE Design \& Test magazine.
\end{IEEEbiography}

\begin{IEEEbiography}[{\includegraphics[width=1in,height=1.25in,clip,keepaspectratio]{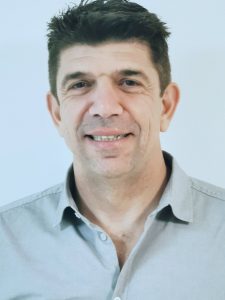}}]%
{Christos Panayiotou} is a Professor with the Electrical and Computer Engineering (ECE) Department at the University of Cyprus (UCY). He is also the Deputy Director of the KIOS Research and Innovation Center of Excellence for which he is also a founding member.  Christos has received a B.Sc. and a Ph.D. degree in Electrical and Computer Engineering from the University of Massachusetts at Amherst, in 1994 and 1999 respectively. He also received an MBA from the Isenberg School of Management, at the aforementioned university in 1999. Before joining the University of Cyprus in 2002, he was a Research Associate at the Center for Information and System Engineering (CISE) and the Manufacturing Engineering Department at Boston University (1999 - 2002). His research interests include modeling, control, optimization and performance evaluation of discrete event and hybrid systems, intelligent transportation networks, cyber-physical systems, event detection and localization, fault diagnosis, wireless, ad hoc and sensor networks, smart camera networks, resource allocation, and intelligent buildings.
    
Christos has published more than 270 papers in international refereed journals and conferences and is the recipient of the 2014 Best Paper Award for the journal Building and Environment (Elsevier).  He is an Associate Editor for the IEEE Transactions of Intelligent Transportation Systems, the Conference Editorial Board of the IEEE Control Systems Society, the Journal of Discrete Event Dynamical Systems and the European Journal of Control. During 2016-2020 he was Associate Editor of the IEEE Transactions on Control Systems Technology. He held several positions in organizing committees and technical program committees of numerous international conferences, including General Chair of the 23rd European Working Group on Transportation (EWGT2020), and General Co-Chair of the 2018 European Control Conference (ECC2018). He has also served as Chair of various subcommittees of the Education Committee of the IEEE Computational Intelligence Society.
\end{IEEEbiography}

\begin{IEEEbiography}[{\includegraphics[width=1in,height=1.25in,clip,keepaspectratio]{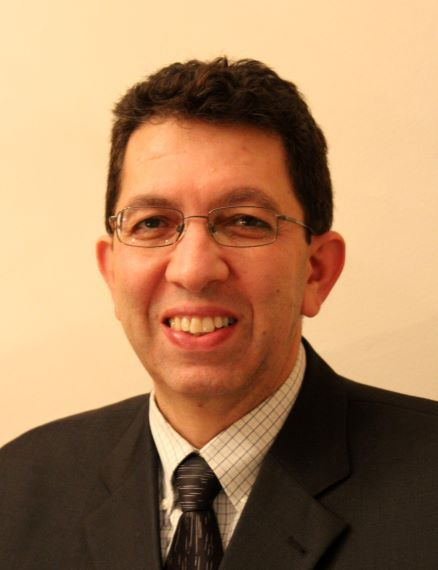}}]%
{Marios M. Polycarpou} is a Professor of Electrical and Computer Engineering and the Director of the KIOS Research and Innovation Center of Excellence at the University of Cyprus. He is also a Member of the Cyprus Academy of Sciences, Letters, and Arts, and an Honorary Professor of Imperial College London. He received the B.A degree in Computer Science and the B.Sc. in Electrical Engineering, both from Rice University, USA in 1987, and the M.S. and Ph.D. degrees in Electrical Engineering from the University of Southern California, in 1989 and 1992 respectively. His teaching and research interests are in intelligent systems and networks, adaptive and learning control systems, fault diagnosis, machine learning, and critical infrastructure systems. Dr. Polycarpou has published more than 350 articles in refereed journals, edited books and refereed conference proceedings, and co-authored 7 books. He is also the holder of 6 patents. 

Prof. Polycarpou received the 2016 IEEE Neural Networks Pioneer Award. He is a Fellow of IEEE and IFAC and the recipient of the 2014 Best Paper Award for the journal Building and Environment (Elsevier). He served as the President of the IEEE Computational Intelligence Society (2012-2013), as the President of the European Control Association (2017-2019), and as the Editor-in-Chief of the IEEE Transactions on Neural Networks and Learning Systems (2004-2010). Prof. Polycarpou serves on the Editorial Boards of the Proceedings of the IEEE, the Annual Reviews in Control, and the Foundations and Trends in Systems and Control. His research work has been funded by several agencies and industry in Europe and the United States, including the prestigious European Research Council (ERC) Advanced Grant, the ERC Synergy Grant and the EU Teaming project.
\end{IEEEbiography}